%% file: Thesis_Upendra.tex
\documentclass[twoside, 12pt]{iiser-thesis}

\usepackage{fullpage}
\usepackage{pdfpages}
\usepackage{amssymb,latexsym,amsmath}     
\usepackage{graphicx}
\usepackage{authblk}
\usepackage{amsthm}
\usepackage{color}
\usepackage{float}
\usepackage{amssymb,amsthm,amsmath}
\newtheorem{definition}{Definition}[section]
\newtheorem{theorem}{Theorem}[section]
\newtheorem{lemma}[theorem]{Lemma}

\newtheorem{corollary}[theorem]{Corollary}
\newtheorem{problem}[theorem]{Problem}
\newtheorem{remark}{Remark}[section]
\usepackage{enumitem}
\usepackage{multirow}
\usepackage[english]{babel}
\usepackage[utf8]{inputenc}
\newcommand\tab[1][1cm]{\hspace*{#1}}
\usepackage{algorithm}
\usepackage{algpseudocode}
\newcommand\Tstrut{\rule{0pt}{2.6ex}}       
\newcommand\Bstrut{\rule[-0.9ex]{0pt}{0pt}} 
\newcommand{\TBstrut}{\Tstrut\Bstrut} 
\newcommand{\etal}{\textit{et al}.}

\newcommand{\N}{Niederreiter }

\catcode`\@=11
\catcode`\@=12

\title{\textbf{McEliece-type Cryptosystems Over Quasi-Cyclic Codes}}

\author{\textbf{Upendra Kapshikar}}
\coordinator{Coordinator} \supervisor{Dr. Ayan Mahalanobis} \sdesignation{ Assistant Professor}
\department{Department of Mathematics} \reader{Dr. Krishna Kaipa}
\dedication{\section*{ }\includegraphics[scale=0.86,page=1]{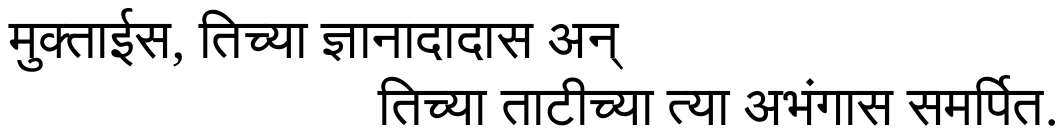}}
\graduationyear{2018}
\academicyear{2017-2018}
\graduationmonth{May}

\thesisabstract{\input{Abstract}}

\acknowledgments{\input{Acknowledgements}}

\begin{document}	
	\thesisfront
\chapter{\textsc{Coding Theory and Public Key Cryptosystems}}
\input{Chapter1}

\chapter{\textsc{Quantum Fourier Sampling}}
\input{Chapter2}

\chapter{\textsc{A McEliece Variant}}
\input{Chapter3}
\chapter{\textsc{Quantum Secure Niederreiter Variant}}
\input{Chapter4}

\chapter{\textsc{Results and Conclusion}}
\input{Chapter5}
\bibliographystyle{ieeetr}
\bibliography{paper}
\end{document}

%% file: Abstract.tex

In 1994, Peter Shor came up with a quantum algorithm that can factor integers efficiently. Along with the world of computations, this revolutionised the field of cryptography as one of the most popular protocols, RSA, assumes hardness of this problem. Though no practical quantum computers existed by then, it was evident that we are in dire need of a cryptosystem which will remain secure even in the era of quantum computers.

Interestingly, Shor's algorithm can be generalised, in a loose sense, to a higher class of problems commonly refereed to as hidden subgroup problems. The hidden subgroup problem can be stated as: given a function $f$ over a group $\mathrm{G}$ that separates cosets \footnote{A fuction that is constant on a coset and different on different cosets} of an unknown subgroup $\textrm{H}$, the task is to find a generating set for subgroup $\mathrm{H}$ in $O(poly(log(\mathrm{G})))$ many measurements and post-processing steps. The class consists of a wide range of problems from order finding ($\mathbb{Z}/ n \mathbb{Z}$) to graph isomorphism ($S_{n}$) and shotest vector problem in lattices ($D_{2n}$). One of the key ingradients of Shor's algorithm is quantum Fourier sampling (QFS). The succuess of the algorithm depends on how effective quantum Fourier sampling is in the corresponding case. The algorithm uses a quantum gate known as QFT (quantum Fourier transform) which performs the natural analogue of discrete Fourier transform over a group. Using this QFT, the quantum Fourier sampling produces a probabilty distribution over irreducible representations of group $\mathrm{G}$  (or on positions inside a matrix of represntaions if one does strong Fourier sampling) based on cosets of $f$. Basic idea to solve a hidden subgroup problem is reconstructing subgroup from obtained probability distribution. This process is feasible in every finite abelian group but the problem is open to a large extent for non-commutative groups. Naturally, this motivates us to look into cryptosystems where the underlying structure is non-commutative.

McEliece cryptosystem, developed by Robert McEliece in 1978, is one such cryptosystem where the underlying structure is non-commutative. McEliece cryptosystem is based on linear algebraic codes, one of the two most promissing enviromnents for post-quantum cryptography (other one being lattice based cryptography). The main advantages for McEliece system other than possibly being quantum secure is faster speed than RSA and ElGamal. Hence, there is a lot of interest in McEliece cryptosystem. However, most of the claim of security does not have a proper theoretical backing. The idea to come up with a proof of security for a McEliece cryptosystem was started by Dinh, Moore and Russel. The idea used by them can be traced back to the work by Julia Kempe and Aner Shalev where they charcterized hidden subgroups of $S_{n}$ that are distinguishable from the identity subgroup. When two subgroups give probabilty distribution from QFS that are \emph{very close } to each other then we can not differentiate between those two and hence, hidden subgroup problem can not be solved. Dinh, Moore and Russel extended this idea for identity subgroup and conjugate subgroups in the corresponding group for McEliece cryptosystem with goppa codes.

Though Niederreiter cryptosytem with goppa codes is shown to be quantum secure by Dinh, Moore and Russel, the system developed by goppa codes has two major drawbacks. One, the key sizes are too large and two, transmission rate is small, approximately 0.52. We extend this result to a new class of Quasi-cyclic codes of a certain kind. We show that it is impossible that the known quantum attack, using hidden subgroup problem and quantum Fourier sampling (or QFS) will break Niederreiter cryptosystem using a particular type of parity check matrices. We use a result provided by Dinh, Moore and Russel for this. To our knowledge, this is only the second variant of McEliece, after classical goppa codes, that is quantum secure.

Our proposed system has a better transmission rate than the previous variant. In most cases, key sizes are significantly smaller than the original variant. We also provide an algorithm that generates a parity check matrices satisfying the required set of conditions.

Chapter 1 gives brief summary of definitons and concepts from coding theory and cryptosystems based on algebraic coding theory. Quantum Fourier sampling and problems of hidden subgroup, hidden shift are reviewed in chapter 2. In chapter 3, we look into a closely related problem of combinatorial optimizaion where we are trying to find a class of QCCs that have small automorrphism group and large enough minimal degree. For this probelm also, we give a class of such codes and provide an algorithm that gives generator matrices for such codes. Chapter 4 provides a Niederreiter variant that is both classically and quantum secure.

%% file: Chapter1.tex
In 1978, Robert McEliece ~\cite{McEliece} came up with a cryptosystem based on algebraic coding theory. The system never gained much popularity due to its large key sizes. Despite of this major drawback, McEliece and similar cryptosystems have started to gain mathematicians' attention as it is believed to be quantum secure. Unlike RSA or ElGamal, McEliece cryptosystem is based on non-commutative structure of algebraic codes. Compared to traditional cryptosystems, McEliece cryptosystem has following advantages
\begin{description}
\item[a)] It is fast. Faster than RSA or ElGamal.
\item[b)] It is believed to be quantum secure.
\end{description}
The major problem with the McEliece cryptosystem is its large key size which makes implementation difficult.
\begin{description}
\item[a)] Key sizes huge.
\item[b)] Cipher-text becomes much larger than the plain-text because of the redundancy added by the encoding process.
\end{description} 

In this chapter, we start with basic definition from coding theory. Then we describe encryption and decryption processes for both McEliece and Niederreiter cryptosystems. Some important classical attacks are reviewed and current specifications of parameters are given.  

\section{Coding Theory}
Error correcting codes are broadly separated in two sections; linear codes and non linear codes. Due to their elegant mathematical structure, linear codes received major interest of the coding community. Traditionally, linear codes are categorized into two types; block linear codes and convolutional codes. As every McEliece cryptosystem is based on block linear codes, we look specifically into block linear codes. Our standard reference for coding theory will be Blahut ~\cite[Chapter 3]{Blahut}. Most of our discussions are on binary linear codes, that is, codes over $\mathbb{F}_{{2}^{l}}$ for some integer $l\geq 1$ but the same theory can be applied over any finite field.

A $q$-ary linear code $\mathcal{C}$ of length $n$ and rank $k$ is a $k$ dimensional linear subspace of $\mathbb{F}_{q}^{n}$

\begin{definition} Hamming weight: Hamming weight or simply weight $w$ of a vector $v$, denoted by $w(v)$, is defined as the number of non-zero entries in $v$.
\end{definition}

The hamming weight sets up a canonical distance function over $\mathbb{F}_{{q}^{n}}$. This distance is called the hamming distance.

\begin{definition} Hamming distance $d_{H}(x,y) :=  w(x-y)$. 

\noindent The distance of code $\mathcal{C}$ is defined as the minimum of the distance between any two distinct codewords in $\mathcal{C}$. We denote this by $d(\mathcal{C})$ or simply $d$ if code $\mathcal{C}$ is clear from the context. 

\[d(\mathcal{C})= \min_{x,y \in \mathcal{C}, x\neq y} d_{H}(x,y)\]. 
\end{definition}

Traditionally, such a code is denoted as $\left[ n,k,d\right]$
Following is an easy lemma, proof of which can be found in any standard text such as ~\cite{Roth,Huffman-Pless}.
\begin{lemma} Let $\mathcal{C}$ be a linear code then \[d(\mathcal{C}) = \min_{x \in \mathcal{C} , x \neq 0} w(x)\]
\end{lemma}

One of the important parameters of a code is its error correction capacity.
\begin{definition} Error correction capacity $t$ : Let $\mathcal{C}$ be $\left[n,k,d \right]$ code. Then the error correcting capacity is the largest integer $t$ such that for any $y\in \mathbb{F}_{q}^{n}$ there exists a unique $c \in \mathcal{C}$ such that $d_{H}(y,c) \leq t$. The ratio $\frac{t}{n}$ is known as the error correction rate of the code $\mathcal{C}$.   
\end{definition}
Error correction capacity and the minimum distance of a code are related to each other by an obvious relation $t= \left\lfloor \dfrac{d(\mathcal{C})-1}{2} \right\rfloor$.
\vspace{\baselineskip} 

A natural way to construct this $k$ dimensional space $\mathcal{C}$ is by viewing it as a range space of some $k \times n$ matrix $G$ known as the generator matrix of the code $\mathcal{C}$ and we say that $G$ generates code $\mathcal{C}$. Alternatively, $\mathcal{C}$ can be constructed by kernel space of some matrix $H$ of size $(n-k) \times n$. In such cases $H$ is known as the parity check matrix. Of course, for a given code $\mathcal{C}$ there are many generator matrices and there are many parity check matrices. Code generated by $H$ is known as the dual code of $\mathcal{C}$ and is denoted as $\mathcal{C}^{\perp}$. Note that both $G$ and $H$ are full rank matrices.

It is easy to check that $GH^{T} = 0 $. Also dual of the dual is the same code,that is, $\left( \mathcal{C}^{\perp} \right) ^{\perp} = \mathcal{C}$. So another way to define dual code is as follows:
\begin{definition} Dual code $\mathcal{C}^{\perp}$ : Let $\mathcal{C}$ be a linear $\left[ n,k \right]$ code over $\mathbb{F}_{q}$ then its $\left[ n, n-k \right]$ dual $\mathcal{C}^{\perp}$ is defined as 

$\mathcal{C}^{\perp} = \lbrace y \in \mathbb{F}_{q}^{n} \text{ such that for all } x\in \mathcal{C}\; \langle x, y \rangle = 0 \rbrace$ where $\langle x,y\rangle$ denotes inner product of $x$ and $y$ over $\mathbb{F}_{q}$.
\end{definition}

Now we give a few examples of codes. 

\begin{enumerate}
\item[$\left(I \right)$] Hadamard Code $\left[ n=2^{r}, k=r, d=2^{r-1} \right]$ :

\noindent Hadamard Code is a linear code generated from a generator matrix whose $i^{th}$ column is the number $i$ written as a binary expression in $r$ bits. Thus, there are $2^{r}$ columns corresponding to all binary numbers with $r$ bits. The code has minimum distance $d=2^{r-1}$ and hence can correct roughly about $t=2^{r-2}$ errors.
\vspace{\baselineskip}

eg. $G=\left[ \begin{array}{cccccccc}
0 &0 &0 &0 &1 &1 &1 &1\\
0 &0 &1 &1 &0 &0 &1 &1\\
0 &1 &0 &1 &0 &1 &0 &1 \end{array} \right] $

\item[$\left(II \right)$] Binary Goppa Codes $\left[n,k,2t+1 \right]$ :

\noindent Construction of a binary Goppa code is done with a polynomial $g(x)$ of degree $t$ over a finite field   $\mathbb{F}_{2^{m}}$ of characteristic 2 without multiple zeros. Then construction of binary Goppa codes is done in the following way

\noindent Pick a set of $n$-points $\lbrace p_{1},p_{2},\ldots ,p_{n}: p_{i} \in \mathbb{F}_{2^{m}} , g(p_{i}) \neq 0 \rbrace$ i.e. none of the $p_{i}'s$ are a root of the polynomial $g(x)$. Then parity check matrix for corresponding code is a product of two matrices; first one is a vandermonde matrix and the second one is a diagonal matrix. Construct following matrices $V$ and $D$
\vspace{\baselineskip}

$V=\left[ \begin{array}{cccc}
 1&1&\cdots& 1\\
 p_{1}&p_{2}&\cdots&p_{n}\\
 p_{1}^{2}&p_{2}^{2}&\cdots&p_{n}^{2}\\
 \vdots& \vdots&\ddots& \vdots\\
 p_{1}^{t-1}&p_{2}^{t-1}&\cdots&p_{n}^{t-1}
\end{array}\right]$ 
$D=\left[ \begin{array}{cccc}
\frac{1}{g(p_{1})}&0&\cdots& 0\\
0& \frac{1}{g(p_{2})}&\cdots&0\\
\vdots& \vdots&\ddots& \vdots\\
0&0&\cdots&\frac{1}{g(p_{n})}
\end{array}\right]$
%

$H = V D $.

\vspace{\baselineskip}
Binary Goppa codes are the codes used in the original McEliece cryptosystems.

\item[$\left(III \right)$] Cyclic Codes:

\noindent Let $a=(a_{1},a_{2},\ldots ,a_{n})$ be an element of $\mathbb{F}_{q}^n$ then we define the right shift of $a$ as $(a_{n},a_{1},\ldots ,a_{n-1})$. A linear code is called a cyclic code if it is closed under right shifts.
 
\noindent Cyclic codes have a nice algebraic structure. The code can be considered as an ideal in the polynomial ring $R=\mathbb{F}_{q}[x]/(x^{n}-1)$ which is a PID. In this ring multiplication by $x$ to a codeword $c$ results into a right shift of $c$. Being an ideal, the code space can be generated by a single polynomial $g(x)$, known as generator polynomial of the code.

\item[$\left(IV \right)$] Quasi Cyclic Codes (QCCs) $\left[ n=m_{1}p, k=m_{2}p, d \right]$: 
Quasi-cyclic code is a simple generalization of a cyclic code. A linear code is considered from a class of quasi-cyclic codes if there exists an integer $m$ such that code is closed under $m$ right shifts. That is, if we denote $R$ as a right shift operator then for all $c \in QCC$ we have $R^{n}(c)\in QCC$. Such lowest $m$ for which the code is closed under $m$ right shifts is known as the index of the code. For cyclic codes, we have $m=1$. QCCs can be constructed by generator matrices whose every block is a circulant matrix of size $p$ and the number of such blocks in a particular row indicates its index. Similar to cyclic code , QCCs also have a nice algebraic structure. We will come back to QCCs later to talk in detail as our both variants are built over QCCs.   
\end{enumerate}
Consider a communication where we want a $k$-bit message $m=(b_{1},b_{2},\ldots,b_{k})$. Now when we send this over a channel, there is a possibility that some $b_{i}$ gets corrupted to $b'_{i}$. Notice that such corrupted vector is also a possible message that one could intend to send. Hence the receiver receives a wrong message. To overcome this difficulty, we add redundancy to the message eg. repeating the message bits thrice, i.e, for message $m=(b_{1},b_{2},\ldots,b_{k})$ of k-bits we instead send a message of $3k$-bits $m'=(b_{1},b_{1},b_{1},b_{2},b_{2},b_{2},\ldots,b_{k},b_{k},b_{k})$. Such codes are known as repetition codes. So even if some bit $b_{i}$ gets corrupted to $b'_{i}$ at one particular position, we have two other copies to recover the original message $m$. \footnote{Note that corrupted message is not in the possible sample set for receiver since theoretically receiver expects vectors with some particular pattern, namely each partition of 3-components has same element throughout the partition.} In practice, much larger numbers than 3 are used. It is highly unlikely that the exact same corruption happens at multiple places.

This process of adding redundancies before sending a message is known as the encoding of an error correcting code. The process for a general linear error correcting code can be implemented as follows: Suppose we want to send messages of $k$-bits. So we have a message space of dimension $k$ over $\mathbb{F}_{q}$. We embed this space into much higher dimensional space, say of dimension $n$, over $\mathbb{F}_{q}$. One of the ways this can be done is by using a compatible linear map $G$ of full rank. 
\begin{center}
$\begin{array}{rccl}
\varphi_{G}: & \mathbb{F}_{q}^{k} &\longrightarrow& \mathbb{F}_{q}^{n}\\
			& x & \mapsto & xG
\end{array}$
\end{center}

The map has a $k$-dimensional range with trivial kernel. As one might guess, our generator matrix $G$ can be used as a linear map. The domain of the map $\varphi_{G}$ is known as the message space while the range is known as the code space. This process is known as encoding and the map $\varphi_{G}$ is known as the encoding function.

Once received, the vector from $\mathbb{F}_{q}^{n}$ has to be converted back to the original message vector in $\mathbb{F}_{q}^{k}$. This is done by the process known as decoding. A decoding process is a two step process. Usually, the first step consists of finding the nearest codeword to the received vector. And the second step is obtaining the message from this codeword. The second step can be done using common routines from linear algebra such as Gaussian elimination and generally is not a difficult task. 

\[\begin{array}{rcccl}
\psi: & \mathbb{F}_{q}^{n} & \stackrel{\psi_{1}}{\longrightarrow} & \mathcal{C} \stackrel{\psi_{2}}{\longrightarrow} & \mathbb{F}_{q}^{k}
\end{array}\]
\begin{center} 
where $\psi_{1}(x) \mapsto c$ such that for all $c^{\prime}\in \mathcal{C},\; d_{H}(x,c^\prime) \geqslant d_{H}(x,c)$ 

and $\psi_{2} $ is an inverse of $\varphi_{G}$ when restricted on $\mathcal{C}$. 
\end{center}
Here, $e=x-c$ is known as the error and process of computing $c$ from $x$, denoted by $\psi_{1}$ above is called the error correction. As this is the hardest computation in $\psi$, sometimes, $\psi_{1}$ is also referred as decoding of a linear code. This process of decoding is not easy for a random linear code. The problem of decoding a random linear code is known as the general decoding problem and is NP-complete ~\cite{brl}. But there are some codes for which this can be done easily eg. RS codes, Binary Goppa codes, cyclic codes, some subclasses of QCCs. Such codes are called decodable codes and it is said that they have a decoder.  

\section{Public Key cryptosystems based on coding theory}

Using this hardness of the general decoding problem, Robert McEliece ~\cite{McEliece} proposed a cryptosystem based on algebraic coding theory. Later Niederreiter proposed his knapsack like variant based on the same principle. The rough idea behind both the systems can be explained as below :

Take a code $\mathcal{C}$ which has a good decoding algorithm. Transform this code into a new code $\mathcal{C^\prime}$ with no apparent structure. A sender sends message using code $\mathcal{C^\prime}$. Now as the code $\mathcal{C^\prime}$ has no visible structure it is as good as a random linear code and no one can decode it. As you are the one who generated the system you have the transformation from $\mathcal{C}$ to $\mathcal{C^\prime}$ you can revert the transformation and work in the frame of $\mathcal{C}$ and then decode successfully as $\mathcal{C}$ has a good decoder. Again, an eavesdropper can not decode this as he does not have a way to transform the system from $\mathcal{C^\prime}$ to $\mathcal{C}$. 

Now we move into McEliece and Niederreiter cryptosystems. We first explain there encryption and decryption algorithms and classical security. We address the question of quantum security in the chapter 2. 
\subsection{Description of the McEliece cryptosystem}
Let $M$ be a generator matrix for a $\left[n,k\right]$ linear code $\mathcal{C}$ for which a fast decoding algorithm exists. Let $\mathcal{E}$ be the number of errors that $\mathcal{C}$ can correct.  

\textbf{Private Key}: ($S$,$M$,$P$) where $S \in GL_{k}(\mathbb{F}_{2})$ and  $P$ is a $n\times n$ permutation matrix.

\textbf{Public Key}: $M^\prime=SMP$.

\noindent\textbf{Encryption}: 
\begin{description}
\item Let $p \in \mathbb{F}_{q}^{k}$ be a $k$-bit plain-text. Corresponding cipher-text $c \in \mathbb{F}_{q}^{n}$ is obtained by calculating $c=pM^\prime+e$ where $e$ is a random error vector such that $wt(e) \leq \mathcal{E}$.
\end{description}
\textbf{Decryption}:
\begin{description}
\item Received cipher-text $c$ is decrypted in the following way:
\item Multiplying $c$ by $P^{-1}$ we obtain $c P^{-1}=(pM^\prime+e)P^{-1}=(pSMP+e)P^{-1}=pSM+e_{2}$. Note that $e_{2}$ has same weight as $e$.

\item Now use the decoding algorithm for $M$ on vector $SAM+e_{2}$ to obtain $pS$.

\item Multiply by $S^{-1}$ to recover plain-text $p$.
\end{description}

\textit{A brief explanation for security}:

Consider a communication where Alice wants to send a message to Bob. Similar to any public key cryptosystem, Bob generates his private key and computes its public counterpart. Private key consists of three matrices $S, M, P $ with $M$ having a good decoder. The private key has a corresponding public key $M^\prime$. Since our $S\; and\; P$ are chosen randomly, the resulting matrix $M^\prime$ is also a random matrix with no structure and hence no efficient decoding with $M^\prime$ is possible. Clearly, every process in decryption is trivial for Bob and he can decrypt easily. An eavesdropper, however, can not do this because he has no good decoding algorithm for publicly known matrix $M^ \prime$ and he does not have knowledge of $S$ and $P$ which is essential.

Another interesting question regrading security of the system is, \textsl{'what happens if $M$  is known ?'} The problem is same as finding transformation $\left( S, P \right)$ between two codes. It is believed that this problem is not easy to solve. Decision version of this, that is, finding if two codes are equivalent is NP-hard and graph isomorphism problem can be reduced to it; so, if one solves the code equivalence problem, one of the problems that received the most attention in last few decades by theoretical computer science community can be solved efficiently ~\cite{Petrank}. This problem of finding transformation between codes or generator matrices of equivalent codes remains intractable. But this presents a possible window of attack from a quantum computer as this transformation problem, known as scrambler-permutation problem, can be modelled as a hidden shift problem and further can be reduced to a hidden subgroup problem where quantum Fourier sampling can come into play. Thus, it becomes essential to make sure that McEliece or its variants resist this quantum Fourier sampling attack.   

\subsection{Niederreiter Cryptosystem}
Let $H$ be a $(n-k) \times n$ parity matrix for a [n,k] linear code $\mathcal{C}$ for which a fast decoding algorithm exists. Let $\mathcal{E}$ be the number of errors that $\mathcal{C}$ can correct.  

\textbf{Private Key}: ($S$,$H$,$P$) where $S \in GL_{k}(\mathbb{F}_{2})$ and  $P$ is a permutation matrix of size $n$.

\textbf{Public Key}: $H^{\prime}=SHP$.

\noindent\textbf{Encryption}: 
\begin{description}
\item Let $p$ be a $n$-bit plain-text with weight at most $\mathcal{E}$. Corresponding cipher-text $c$ of $n-k$ bits is obtained by calculating $c=H^\prime p^T$.
\end{description}
\textbf{Decryption}:
\begin{description}
\item Compute $y=S^{-1} c$. Thus $y = H P {{p}^T}$.
\item By linear algebra find a $z$ such that $Hz^T = y$. As $y = H P {{p}^T}$ we have $z-pP^T \in \mathcal{C}$.
\item Now use fast decoding on $z$ with $H$ to get $pP^T$ and thus recover $p$.\end{description}
\subsection{Signature scheme}
Initially it was thought that McEliece cryptosystem could not accommodate for signature scheme as it is not a commutative cryptosystem in the sense that order or role of encryption and decryption algorithms can not be altered. In other cryptosystems such as RSA, where encryption and decryption algorithms do commute, ready signature schemes are available. Later, in 2001, Courtois, Finiasz and Sendrier ~\cite{Courtois} came up with a signature scheme for Niederreiter cryptosystem.

\section{Classical attacks}
In this section we briefly go over the generic classical attacks. Most of the attacks are local attacks in the sense that they try to decrypt a given ciphertext. Complete breaks that completely recover private keys are extremely demanding and not possible.

The attacks trying to recover plaintext from the knowledge of ciphertext and public key are really hard due to the general decoding problem and are out of the discussion. Most of the classical attacks that stand a chance come under category called Information Set Decoding (ISD). There are two popular ISD attacks; one by Stern ~\cite{Stern} and other by Lee and Brickell ~\cite{Lee}. As mentioned in ~\cite{Baldi}, ISD attacks are the best known classical attacks and hence considered as security level of the system. 

One of the basic attacks was suggested by McEliece~\cite{McEliece}. Lee and Brickell improved his attack and added an important verification step where attacker can confirm whether recovered message is the correct one. The strategy is based on repeatedly selecting $k$ bits at random from an $n$-bit cipher-text in hope that none of the selected bits are part of the error. Similar attacks can also be implemented over Niederreiter cryptosystems. Lee and Brickell also provided a closed-form equation for complexity of the attack. The work factor for this attack can be given as
\vspace{\baselineskip}

$W_{j} = T_{j} \left( \alpha k^{3} + N_{j} \beta k \right)$ where, 
\vspace{\baselineskip}

\tab[3cm] $T_{j} = \dfrac{1}{\sum\limits_{i=0}^{i=j} \dfrac{{t \choose i} {n-t \choose k-i}}{{n \choose k}}} $  
\tab[1cm]  and \tab $N_{j} = \sum\limits_{i=o}^{i=j}{k \choose i}$
\vspace{\baselineskip}

Other attack is by Stern ~\cite{Stern}. The basic idea behind this attack is to recover intentional vector by embedding public code space into some higher dimensional codespace. Let $C^\prime$ be the code generated by public key $G^\prime$ then Stern constructs a code given by generator matrix $G^{{\prime}{\prime}}= \left[ \begin{array}{c}
G^{\prime} \\
x
\end{array} \right]$ . Bernstein, Lange and Peters ~\cite{Bernstein} improved the attack using Markov Chain Modelling. This modified version of attack made the attack faster by a factor of $2^{12}$ and parameters for the system had to be readjusted.

For a given code, both McEliece and Niederreiter cryptosystem have same security ~\cite{Li} on the equation level. That means, if one tries to recover plain-text from cipher-text for a McEliece cryptosystem it is hard as doing the same for Niederreiter cryptosystem and hence ISD attacks have same strength when applied on McEliece or Niederreiter over a code with same parameters. Li, Deng and Wang ~\cite{Li} also analyzed both the systems under the attack by Lee and Brickell.  

\subsection{Parameters}
Robert McEliece in his original work suggested parameters $n = 1024,\; k=524,\; t=50$. As mentioned before, after ~\cite{Bernstein} these parameters were not secure. Bernstein suggested some new sets of parameters. We mention a few of them here.
\begin{description}
\item[(a)] For Goppa codes and 80 bit security $n=1632$, $k=1269$, $t=33$. The public key size in this case is $460647\; bits$.
\item[(b)] Without list decoding suggested set is $n=2048,\; k=1751,\; t=27$. The public key size in this case is $520047\; bits$
\item[(c)] For 128-bit security $n=2960$, $k=2288$, $t=56$. This parameter selection leads to public keys of size $1537536$ $bits$. 
\end{description} 
From these parameter choices it is very clear that McEliece cryptosystem or its Niederreiter variant suffers two major drawbacks:

\begin{description}
\item[$(I)$] Large public key sizes.
\item[$(II)$] Low transmission rate or encryption rate.
\end{description} 

Various attempts have been made to overcome above problems but most of them turned out to be insecure. McEliece variants over Quasi-cyclic circulant codes ~\cite{Baldi} is one of the notable attempts in that direction. In this particular version, authors looked at QC-LDPC codes and put forth a cryptosystem which is similar to the McEliece cryptosystem with matrices $S$ and $P$ replaced by block circulant matrices. Though in this variant shorter key structure is possible and higher encryption rates can be achieved, one of the most important piece of the puzzle, quantum security of this variant is still missing. It is absolutely essential that if McEliece or its any variant were to replace any of the current popular systems such as RSA, ElGamal first priority should be its quantum security than key sizes and encryption rate. 

In next few chapters we will look into quantum security of McEliece-type cryptosystems and then provide a Niederreiter variant that is quantum secure having high transmission rate. We also present a brief comparative analysis of key sizes and encryption rate to original McEliece cryptosystems.

%% file: Chapter2.tex
In this chapter, we look at basics of quantum Fourier sampling. We begin by defining a problem in abelian group theory.

\begin{problem} Let $f$ be a function from $\mathbb{Z}/N\mathbb{Z}$ to $\mathbb{Z}/N\mathbb{Z}$ defined as \[\begin{array}{rcllr}
f(x): &  \mathbb{Z}/N\mathbb{Z} & \longmapsto & \mathbb{Z}/N\mathbb{Z} & \multirow{2}{*}{for some constant $a \in \left(\mathbb{Z}/N\mathbb{Z}\right)^{\times}$} \\
	& x & \mapsto & a^{x}
\end{array}.\] We want to find period the of $f(x)$.
\end{problem}
Classically, it is well known that this problem is very hard. Integer factorization problem can be reduced to above order finding problem and hence one can solve integer factorization problem by solving order finding problem. However, with quantum computer this problem can be efficiently solved. The algorithm that solves this problem was given by Shor ~\cite{Shor} ~\cite{Shor2} in 1995. This algorithm is the first quantum algorithm for some practical problem. Before Shor's algorithm, there were a few interesting algorithms that showed glimpses of the power of a quantum computer ~\cite{Deutsh} ~\cite{Simon} but these problems have artificial flavor in them and have hardly little practical importance.

Shor's algorithm was a great boost to algorithmic computer science in itself. Along with solving an important question for cyclic groups this algorithm opened doors for much powerful mathematics. Naturally, this situation presented two questions in front of computer science community. One, can we solve this order finding problem in a more general setting, say an abelian group or more generally in any group and two, if we were to extend this order finding function to some class of functions what would be right choice for the problem and will we be able to solve this problem?

Interestingly, both the questions have a common answer, quantum Fourier Sampling (QFS). In some sense, QFS can be viewed as a general form of Shor's algorithm. Quantum Fourier Sampling is an important tool that acts behind the scene for almost all known quantum algorithms that offer exponential speed up compared to classical algorithms. It is the reason why Shor's and Simon's algorithms work. The order finding problem is a special case of what is known as a hidden subgroup problem. 
\section{The Hidden Subgroup Problem (HSP)}
\begin{definition} {Hidden Subgroup Problem}: Let $\mathrm{G}$ be a group and $f$ a function from $\mathrm{G}$ to a set $\mathrm{X}$. We know that  $f(g_{0})=f(g_{1})$ if and only if $g_{0}\mathrm{H}=g_{1}\mathrm{H}$ for some subgroup $\mathrm{H}$. The problem is, given $f$ find a generating set for the unknown subgroup\footnote{The function $f$ in the hidden subgroup problem is said to be separating cosets of $\mathrm{H}$ as $f$ is constant on a each coset and different on different cosets.} $\mathrm{H}$.

\end{definition}
In particular, Shor's algorithm is a hidden subgroup problem over $\mathbb{Z}/N\mathbb{Z}$ with function $f(x)=a^{x}$. In this case hidden subgroup is $\mathrm{H}= \langle r\rangle$ where $r$ is the order of $a$ in $\left( \mathbb{Z}/N\mathbb{Z} \right) ^{\times}$.

Quantum Fourier Sampling is an algorithm which uses Quantum Fourier Transform as a building block. Before going into QFS we recall some of the facts form representation theory. Given a group $\mathrm{G}$ a matrix representation is a homomorphism $\rho: \mathrm{G}$ $\longrightarrow$ $GL_{d_{\rho}}\left(\mathbb{C}\right)$ where $GL_{d}\left(\mathbb{C}\right)$ is a space of $d \times d$ matrices over complex numbers. We denote set of all the irreducible representations of the group $\mathrm{G}$ by $\widehat{\mathrm{G}}$. So for every $\rho \in \widehat{\mathrm{G}}$ and every $g \in \mathrm{G}$; $\rho(g)$ gives us a $d_{\rho}$ $\times d_{\rho}$ matrix and $\rho_{(i,j)}(g)$ will denote the the entry in $i^{th}$ row and $j^{th}$ column of $\rho(g)$. We stick only to finite groups and their complex irreducible representations. One of the very well known result from representation theory of finite groups states that $\sum_{\rho} {d_{\rho}}^{2}= \vert \mathrm{G} \vert $.

Let $\vert \mathrm{G} \vert = N$. Fix an ordering $\left( g_{1},{g}_{2},\ldots,g_{N}\right)$ on elements of $\mathrm{G}$. For a vector in $\mathbb{C}^{N}$, a general normalized state in basis $\mathcal{B}_{1} = \lbrace g_{i};\; 1\leqslant i \leqslant N\rbrace$ can be described as  \[\vert \psi \rangle = \sum_{i=1}^{N} \alpha_{i} \vert g_{i}\rangle\;  \text{where }\alpha_{i}s\; \text{are complex numbers such that} \sum_{i=1}^{N}\vert \alpha_{i}\vert^{2} = 1.\] 

Now consider another basis for $\mathbb{C}^{N}$ given by $\mathcal{B}_{2}= \lbrace (\rho,i,j)\; \rho\in \widehat{\mathrm{G}},\;1 \leqslant i,j \leqslant d_{\rho} \rbrace$. Clearly, $\vert \mathcal{B}_{2}\vert = N$ as $\sum_{\rho} d_{\rho}^{2} = N$. A general normalized state in basis $\mathcal{B}_{2}$ is denoted as 

\[\vert \psi \rangle = \sum_{\rho,i,j} \beta_{\rho,i,j} \vert \rho, i,j \rangle \text{ where }\beta_{\rho,i,j}'s\; \text{are complex numbers such that} \sum_{\rho,i,j} \vert \beta_{\rho,i,j}\vert^{2} = 1.\]
A quantum Fourier transform is a map that takes a normalized state in basis $\mathcal{B}_{1}$ to a normalized state in $\mathcal{B}_{2}$. Under this map the vector associated with $g$ is $\sum_{\rho,i,j}\dfrac{\sqrt{d_{\rho}}}{\sqrt{\vert \mathrm{G} \vert}} \rho_{(i,j)}(g) \vert \rho,i,j \rangle$. Quantum Fourier transform can be viewed as $\mathcal{C}$ linear extension of the above association. Precisely, quantum Fourier transform on a general normalized state $\psi = \sum_{l}\alpha_{l} \vert g_{l} \rangle$ in basis $\mathcal{B}_{1}$ would give 
\[QFT \vert \psi \rangle = \sum_{l} \alpha_{l}\; QFT \vert g_{l}\rangle = \sum_{l} \alpha_{l}  \sum_{\rho,i,j}\dfrac{\sqrt{d_{\rho}}}{\sqrt{\vert \mathrm{G} \vert}} \rho_{(i,j)}(g) \vert \rho,i,j \rangle.\]

Measurement is a nonlinear operator used in almost every quantum algorithm. Measurement operator is defined with respect to a basis. Here we describe measurement with respect to an orthonormal basis only as both $\mathcal{B}_{1}$ and $\mathcal{B}_{2}$ described above form an orthonormal basis. For measurement in a general basis or more general measurement operator see ~\cite{NDavidMermin}, ~\cite{Kaye}. Consider an orthonormal basis for $n$ dimensional space $\mathcal{B}= \lbrace b_1, b_2,\ldots b_n\rbrace$. Then a normalized general state in this basis can be represented by $\vert \psi \rangle = \sum_{i} \beta_{i} \vert b_{i} \rangle$. Measurement on  state $\vert \psi \rangle$ gives $b_{i}$ as its output with probability $\vert \beta_{i}\vert ^{2}$. In other words, after measurement the state collapses to one of the basis element states; the probability that it falls on a particular state depends on coefficient of that basis element in the state before measurement. 

After this short background on quantum computing, we are ready to describe quantum Fourier sampling. For further reading a reader can refer to ~\cite{Kaye,Lomont}

\begin{algorithm}
\caption{Quantum Fourier Sampling (QFS)}\label{alg:QFS}
\begin{algorithmic}[1]
\Procedure{QFS}{$\mathrm{G}, f$}\Comment{QFS over group $\mathrm{G}$ with function $f$}
\State $\vert \psi \rangle = \sum_{g}{\vert g, 0\rangle}$
\State apply $U_{f}$ on $\vert \psi \rangle$ Let $\vert \psi_{2}\rangle = U_{f} \vert \psi \rangle$ 

\Comment{$U_{f}$ is a two state unitary operator such that $U_{f}\vert x,y\rangle = \vert x, y \oplus f(x)\rangle$}
\State Measure in the second component to get $\vert \psi_{3}\rangle$

 \Comment{Measurement of vector $\vert \phi \rangle = \sum_{i} \alpha_{i} \vert e_{i} \rangle$ gives output $e_{i}$ with probability $\vert \alpha_{i} \vert ^{2}$}
\State Apply QFT on first component of $\vert \psi_{3} \rangle$
\State Measure $\vert \rho,i,j\rangle$ for strong Fourier sampling \emph{OR} 

Measure $\vert \rho \rangle$ component for weak Fourier sampling
\EndProcedure
\end{algorithmic}
\end{algorithm}

Strong quantum Fourier sampling gives you output $\vert \rho,i,j \rangle$ for hidden subgroup $\mathrm{H}$ with probability given by 
\[ \mathsf{P}_{\mathrm{H}} \left( \vert \rho,i,j \rangle\right) = \dfrac{1}{\vert \mathrm{G} \vert} \sum_{g\in \mathrm{G}} \mathsf{P}_{g\mathrm{H}} \left( \vert \rho,i,j\rangle \right)\] where  \[\mathsf{P}_{g\mathrm{H}} = \dfrac{d_{\rho}}{\vert \mathrm{G}\vert \vert\mathrm{H}\vert}\left\vert \sum_{h\in \mathrm{H}} \rho_{i,j}  \left(g \mathrm{H} \right)\right\vert ^{2}\]

Details about quantum Fourier sampling and in general status about non-abelian hidden subgroup problem can be found in ~\cite{Vazirani}. For basics of quantum computation and quantum Fourier sampling specifically from hidden subgroup point of view we refer ~\cite{Lomont}. For more broader view of quantum computation reader can use standard texts such as ~\cite{NDavidMermin},~\cite{Kaye}. The rough idea behind using this QFS to solve a hidden subgroup problem is to try and reconstruct $\mathrm{H}$ from $\mathsf{P}_{\mathrm{H}}$. We make this idea precise in next few sections. But before going there, let us briefly look at how hidden subgroup problem can be used to break McEliece cryptosystem or its variants. We define scrambler - permutation problem.
\section{McEliece-type Cryptosystems and HSP}
\begin{problem} {Scrambler - Permutation Problem: } Consider two $k \times n$ matrices $M$ and $M'$ over $\mathbb{F}_{{q}^{l}}$. It is known that they are related by equation $M^{\prime} = SMP$ for some unknown $S \in GL_{k}(\mathbb{F}_{q})$ and some unknown permutation matrix $P$ of size $n$. The problem is to find $S$ and $P$. 
\end{problem}
Clearly, one of the ways to attack McEliece or Niederreiter cryptosystem is by solving the scrambler-permutation problem. For this attack, we assume attacker knows both $M$ and $M'$ and he is trying to recover remaining part of the private key. This attack is known as the scrambler - permutation attack. As far as the quantum attacks go, this is the only known way of attacking a McEliece cryptosystem. This structural attack is exactly same for a McEliece cryptosystem or a Niederreiter cryptosystem except that instead of finding a scrambler-permutation pair from generator matrix $G$ to $G^\prime$ one has to find scrambler-permutation pair from parity check matrix $H$ to $H^\prime$. The algebraic structure of the problem remains the same. So, we present it in general form, to find a scrambler-permutation pair from a $k \times n$ matrix $M$ to other $k \times n$ matrix $M^{\prime}$ keeping in mind $M=G$ and $M^{\prime}= G^{\prime}$ in case of McEliece while $M=H$ and $M^{\prime}=H^{\prime}$ in case of Niederreiter. To mean either a McEliece or a Niederreiter cryptosystem we use a broad term McEliece-type cryptosystems. In this attack, we assume $M$ and $M^\prime$ known, the attack is to find $A$ and $B$ such that $AMB=M^{\prime}$ with $A$ and $B$ coming form groups as defined before. Notice finding  any $A^\prime$ and $B^\prime$ such that $A^\prime MB^\prime=M^\prime$ will also make the attack successful.

\begin{problem}[Hidden Shift Problem] Let $\mathrm{G}$ be a group. Let $f_{0}$ and $f_{1}$ be two functions from group $\mathrm{G}$ to a set $\mathrm{X}$. Given $f_{0}(g)=f_{1}(g_{0}g)$ for some unknown constant $g_{0}$ the task is to find a constant $g_{0} \in \mathrm{G}$. Note that there can be many $g_{0}$ that satisfy the above condition. Hidden shift problem asks us to find any one of those constants.
\end{problem}

Let $M'=AMP$. A McEliece-type cryptosystem will be broken if we find one possible pair $(A,P)$ from $M$ and $M'$. Consider two functions from group $\mathrm{G}=GL_{k}(\mathbb{F}_{2}) \times S_{n} $ given by  
\begin{equation}
f_{0}(A,P)=A^{-1}MP 
\end{equation}
\begin{equation}
f_{1}(A,P)=A^{-1}M'P
\end{equation}
Then one can check that $f_{1}(A,P)=f_{0}((A_{0}^{-1},P_{0}).(A,P))$, that is $(A_{0}^{-1},P_{0})$ is the shift between $f_{0}$ and $f_{1}$. Hence, if one can solve the hidden shift problem over $\mathrm{G}=GL_{k}(\mathbb{F}_{2}) \times S_{n}$ he can break the McEliece-type cryptosystem.

The general procedure to solve this hidden shift problem is to reduce it to try and reduce it to a hidden subgroup problem. We can reduce the hidden shift problem with functions$f_{0}$ and $f_{1}$ defined above on the group $\mathrm{G} =GL_{k} (\mathbb{F}_{2}) \times S_{n}$ to the hidden subgroup problem over $(\mathrm{G}\times \mathrm{G})\rtimes \mathbb{Z}_{2}$ ~\cite[Section 2.2]{Dinh}. The hidden subgroup in this case is

\begin{equation} \label{eqnK}
 \mathrm{K}=(((\mathrm{H}_{0},s^{-1}\mathrm{H}_{0}s),0) \cup ((\mathrm{H}_{0}s,s^{-1}\mathrm{H}_{0}),1))
 \end{equation} where $\mathrm{H}_{0}=\lbrace (A,P)\in GL_{k} (\mathbb{F}_{2}) \times S_{n} : A^{-1}MP=M  \rbrace$ and $s$ is a shift from $f_{0}$ to $f_{1}$.

In short, the scrambler-permutation problem is one of the key ways to attack a McEliece-type cryptosystems. This problem can be formulated as a hidden shift problem which further can be reduced to a hidden subgroup problem. So we can attack McEliece type cryptosystems by trying to solve a hidden subgroup problem over $(\mathrm{G}\times \mathrm{G})\rtimes \mathbb{Z}_{2}$ with 
$\mathrm{G} =GL_{k} (\mathbb{F}_{2}) \times S_{n}$.
\section{Successful Quantum Fourier Sampling}
In the previous section we saw that solving the hidden subgroup problem as a standard way to attack the a McEliece-type cryptosystem. An interesting question is, when is the hidden subgroup problem hard  to solve? This way we can ensure the security of a McEliece-type cryptosystem against known quantum attacks.

We briefly sketch thought behind effectiveness of QFS. Algorithm of QFS in a general scenario and its use for solving a hidden subgroup problem is very well explained in~\cite{Vazirani}. Arguments particular to McEliece-type cryptosystems and corresponding hidden subgroup problem are in~\cite{Dinh}. The standard model of QFS yields a probability distribution as a function of the hidden subgroup. The basic idea here is if two subgroups $\mathrm{H}_{1}$ and $\mathrm{H}_{2}$ yield probability distributions $\mathsf{P}_{\mathrm{H}_{1}}$ and $\mathsf{P}_{\mathrm{H}_{2}}$ such that  $\mathsf{P}_{\mathrm{H}_{1}}$ and $\mathsf{P}_{\mathrm{H}_{2}}$ are \emph{very close} to each other then QFS will not give us enough information to solve the hidden subgroup problem. The concept of closeness of two probability distributions can be captured by setting up a norm on the space of probability functions. To our knowledge, J. Kempe and A. Shalev~\cite{Kempe} were the first to introduce this beautiful idea. They used total variation norm. So, if two probability functions have total variation distance between them less than or equal to $log^{-\omega(1)}\vert \mathrm{G} \vert$ then we say that those two distributions are non-distinguishable. Using this definition, they provided a necessary condition to distinguish a subgroup of $S_{n}$ from the trivial subgroup $\langle e \rangle$. 
Later Dinh, Moore and Russel~\cite{Dinh} extended this result with keeping McEliece-type group structure under consideration. Their result can be viewed as an analysis of a hidden subgroup problem over the  group $\mathrm{G}=(GL_{k}(\mathbb{F}_{2})\times S_{n})^{2} \rtimes \mathbb{Z}_{2}$, the group structure for McEliece-type cryptosystems. Instead of using total variation distance, they use $L_{1}$ distance. Other key difference that can be considered between two definitions is Kempe and Shalev defined it for weak Fourier sampling while Dinh, Moore and Russel defined it for strong Fourier sampling. Also to account for all the conjugate subgroups Dinh, Moore and Russel took expectation over all the conjugate subgroups along with expectation over irreducible complex representations of group $\mathrm{G}$ denoted as $\rho$. Here they demonstrate a case when the hidden subgroup $\mathrm{H}$ can not be distinguished from either its conjugate subgroups $g\mathrm{H}g^{-1}$ or the trivial subgroup $\langle e \rangle$.

First  note that weak Fourier sampling gives same distributions for all the conjugate subgroups that is $\mathsf{P}_{\mathrm{H}}$ is same as $\mathsf{P}_{g\mathrm{H}g^{-1}}$. Hence weak Fourier sampling can not differentiate a subgroup from its conjugate subgroup and thus it suffices to look at strong Fourier sampling.
Dinh, Moore and Russel~\cite{Dinh}, inspired from J. Kempe and A. Shalev~\cite{Kempe} define distinguishability of a subgroup $\mathrm{H}$ by strong Fourier sampling.

\begin{definition} Distinguishability of a subgroup on strong QFS 

$\mathcal{D}_{\mathrm{H}} := \mathbf{E}_{\rho ,g} \Vert \mathsf{P}_{g\mathrm{H}g^{-1}} (\cdot| \rho ) - \mathsf{P}_{\langle e \rangle} (\cdot| \rho )\Vert_{1}$

A subgroup $\mathrm{H}$ is called \textit{indistinguishable} by strong Fourier sampling if $\mathcal{D}_{\mathrm{H}} \leq log^{- \omega(1)} \vert \mathrm{G} \vert$.
\end{definition}  
The real $\mathcal{D}_{\mathrm{H}}$ is nothing but the expectation of $L_{1}$ distance between probability distribution of conjugate subgroups and the trivial subgroup.

Note that if a subgroup $\mathrm{H}$ is indistinguishable according to this definition then by Markov's inequality for all $c$, $\Vert \mathsf{P}_{g\mathrm{H}g^{-1}} (\cdot| \rho ) - \mathsf{P}_{\lbrace e \rbrace} (\cdot| \rho )\Vert _{t.v.} \leq log^{-c} \vert \mathrm{G} \vert $; which is analogous to definition provided by J. Kempe and A. Shalev~\cite{Kempe} for indistinguishability of a subgroup by weak Fourier sampling.
 
Now we state a few definitions which will be used to establish quantum security of McEliece-type cryptosystem.
\begin{definition} $Aut(M)=\lbrace P\in S_{n}$ such that  there exists $ A \in GL_{k}(\mathbb{F}_{q}) ,\ AMP=M \rbrace $.
 \end{definition}
\begin{definition} The minimal degree of a $G \leqslant S_{n}$ acting on set of $n$ symbols is defined to be minimum number of elements moved by a non-identity element of the group $G$.
\end{definition}
\begin{definition}Consider a  $k\times n$ matrix $M$, we define $T_M$ for matrix $M= \left[I_{k} | M^{*} \right]$ as 

$T_M=\lbrace \mathcal{P}_{1} \in S_{k}$ such that there exists $\mathcal{P}_{2}\in S_{n-k}$ with $\mathcal{P}_{1} M^{*} \mathcal{P}_{2} =M^{*}$ $\rbrace$.
\end{definition}
\begin{theorem}
~\cite[Theorem 4]{Dinh}: Assume $q^{k^2} \leqslant n^{an}$ for some constant $0 < a < 1/4$. Let $m$ be the minimal degree of the automorphism group $Aut(M)$. Then for sufficiently large n, the subgroup $\mathrm{K}$, $D_{K} \leqslant O(\vert K \vert ^ 2 e^{-\delta m} ) $, where $\delta > 0$ is a constant.
\end{theorem}
In the above theorem, the subgroup $\mathrm{K}$ is the hidden subgroup for McEliece-type cryptosystems that we stated earlier in this chapter. Details of the proof can be found in ~\cite{Dinh}. For a matrix $M$ of full column rank, $\vert K\vert = 2 \vert {Aut(M)} \vert ^ {2}$ ~\cite{Dinh}. Hence if $\vert Aut(M) \vert ^{4} e^{-\delta m} \leq log^{- \omega(1)} \vert \mathrm{G} \vert$ then $K$ is indistinguishable making scrambler-permutation attack using QFS infeasible. Thus if a $k \times n$ matrix $M$ with minimal degree $m$ is such that $\vert Aut(M) \vert ^{4} e^{-\delta m} \leq log^{- \omega(1)} \vert \mathrm{G} \vert$ then we can not find a scrambler-permutation pair and hence system remains secure against quantum Fourier sampling.Later we use this result for parity check matrix $H$ to show that our Niederreiter cryptosystem is secure against this hidden subgroup attack.

%% file: Chapter3.tex
In this chapter, we describe a new variant of McEliece cryptosystem based on quasi-cyclic codes. We then show that these codes have small automorphism group and large minimal degree. Due to a Dinhtheorem by Dinh, Moore and Russell~\cite{Dinh}, that we looked at in the previous chapter, it becomes a natural direction to look for codes having automorphism groups with small size and large minimal degree. We would like to point out that our system is not necessarily quantum secure as it does not follow the require condition for theorem to hold, that is, $q^{k^{2}} \leqslant n^{\alpha n}$ for some $0 < \alpha < 1/4$. Though our system is not necessarily quantum secure, it presents an interesting mathematical problem of combinatorial optimization where we are looking to reduce the automorphism group size and increase the minimal degree. Quasi-cyclic codes represent an important class of block linear codes and particularly, $\frac{m-1}{m}$ quasi-cyclic codes and $\frac{1}{m}$ codes have received some special attention ~\cite{Gulliver_thesis}. So, it becomes an important question if we can set up McEliece variant based on these quasi-cyclic codes satisfying required bounds on automorphism group size and minimal degree. Apart from, mathematical interests, this variant has encryption rate much higher than original McEliece. We show such construction over codes of rate $\frac{m-1}{m}$ and then note that similar construction can be used to construct $\frac{1}{m}$ codes. In the next chapter, we provide a similar construction of a Niederreiter variant which is indeed quantum secure.

\section{Quasi-Cyclic Codes}

\begin{definition} Cyclic code: A code $\mathcal{C}$ is called a cyclic code if it is closed under right shifts i.e. for all $c = \left(c_{0},c_{1},c_{2},\ldots,c_{n-1} \right) \in \mathcal{C}$ we have  $c^\prime = \left( c_{n},c_{0},c_{1},\ldots,c_{n-1} \right) \in \mathcal{C}$.
\end{definition}

A quasi-cyclic code (QCC), a simple generalization of the cyclic code, is such that any cyclic shift of a codeword by $m$ symbols gives another codeword of QCC. If $m=1$ the code is a cyclic code.
We are particularly interested in $\frac{m-1}{m}$ rate codes. More specifically, our system is based on $\frac{m-1}{m}$ rate codes over $ \mathbb{F}_{2}$ in this chapter and over $\mathbb{F}_{{2}^{l}}$ in the next chapter. Such codes along with Quasi-cyclic codes of rate $\frac{1}{m}$ are studied in great detail in ~\cite{Gulliver_thesis}. 

\begin{definition} Circulant matrix: A $p \times p$ matrix $C$ is called circulant if every row, except for the first row, is a circular right shift of the row above that.
\end{definition} 
A typical example of a circulant matrix is 
\begin{center}
$\left[ \begin{array}{cccc}
c_{0} & c_{1} & \cdots & c_{p-1} \\ 
c_{p-1} & c_{0} & \cdots & c_{p-2}\\
\vdots & \vdots & \ddots & \vdots\\
c_{1} & c_{2} & \cdots & c_{0} 
\end{array} \right]$ \end{center}

Now we state a couple of relevant results about circulant matrices and cyclotomic polynomials. Each lemma is easy to check involving basic ring definitions and Chinese remainder theorem for rings so proofs are skipped. 
\begin{lemma} \label{cir1} Let $\mathcal{C}_{n}$ denote class of all circulant matrices of size $n$. The class $\mathcal{C}_{n}$ forms a commutative ring under usual matrix addition and multiplication. 
\end{lemma}
\begin{lemma} The \label{cir2}Ring $\mathcal{C}_{n}$ is isomorphic to the ring $\frac{\mathbb{F}_{q} \left[ x\right]}{\left(x^{n}-1\right)}.$ Furthermore if $n$ is prime and $p \neq q$ then this isomorphism decomposes as \[ \mathcal{C}_{p} \xrightarrow[]{\sim} \dfrac{\mathbb{F}_{q} \left[ x\right]}{\left(x-1\right)} \times \dfrac{\mathbb{F}_{q} \left[ x\right]}{\Phi_{p}(x)}\] where $\Phi_{p} (x)$ is $p^{th}$ cyclotomic polynomial. \end{lemma}

A rate $\frac{m-1}{m}$ systematic Quasi-Cyclic code has an $p \times mp$ parity check matrix of the form 

$H= \left[I_{p}|C_{1}|C_{2} |\cdots | C_{m-1}\right]$ where each $C_{i}$ is a circulant matrix of size $p$ and $I_{p}$ is identity matrix of size $p$. For compactness we denote this as $H= \left[I| C\right]$ with keeping in mind that $C$ is an array of circulants and we denote $C =$\textsc{Array}$\left[C_{1},C_{2},\ldots,C_{m-1}\right]$. Alternatively, these codes can be defined using generator matrices ~\cite{Aylaj}. In this case, generator matrix takes the following form:
\[ G = \left[ \begin{array}{c|c}
\multirow{4}{*}{I} & {C^{\prime}}_{1}\\
& {C^{\prime}}_{2}\\
& \vdots \\
& {C^{\prime}}_{m-1}

\end{array} \right]\] Again for compactness we denote this as $G=\left[I|C\right]$ with understanding that $C$ is a stack of circulants and we denote $C$ as $C=$\textsc{Stack}$\left[{C^{\prime}}_{1},{C^{\prime}}_{2},\ldots,{C^{\prime}}_{m-1}\right]$.

In a recent work ~\cite{Aylaj} a way to generate generator matrices for such codes over $\mathbb{F}_{{2}}$ is presented. Since these generator matrices are in systematic form one can easily construct parity check matrix from generator matrix. Regarding the codes over extension fields ~\cite[chapter 6]{Gulliver_thesis} shows that Quasi-cyclic codes over extension fields can be MDS (maximum distance separable) codes. As the name suggest MDS codes can achieve large minimum distance and hence no low weight codewords. This plays an important role for classical security of the system against classical attacks such as Stern's attack and Lee-Brickell attack. Though ~\cite{Gulliver_thesis} presents examples of MDS codes with rate $\frac{1}{m}$, this does present a case for study quasi-cyclic codes of rate $\frac{m-1}{m}$ with large minimum distance. For more details about quasi-cyclic codes a reader can refer to ~\cite{Gulliver_thesis,Aylaj}.

\subsection{Decoding} Quasi-cyclic codes are well studied and well established codes and depending on how one constructs them there are various decoders available. We briefly mention some of them here. ~\cite[Appendix B]{Gulliver_thesis} presents some ML (majority logic) decodable QCCs. Another new and interesting way of decoding quasi-cyclic codes using Gr\"{o}bner basis formulation can be found in ~\cite{Ling}.

\section{Our McEliece variant}
Now we are ready to describe our McEliece variant over quasi-cyclic codes of rate $\frac{m-1}{m}$. Our McEliece variant over $\mathbb{F}_{2}$ has a generator matrix $M = \left[I|C \right]$ with $C=$\textsc{Stack}$\left[C_{1},C_{2},\ldots,C_{m-1}\right]$ satisfying following conditions:
\begin{description}
\item[$(I)$] Size of each circulant is a prime $p$, i.e., each circulant is a $p \times p$ matrix for some prime $p$.
\item[$(II)$] At least one of the $C_{i}s$ is invertible i.e. there exists $i<m$ suth that $C_{i} \in GL_{p}(\mathbb{F}_{2})$.
\item[$(III)$] Given any two columns $c_{i_{0}}$, $c_{i_{1}}$ of $C$, there is at most one index $j$ with $c_{i_{0}}[j]=c_{i_{1}}[j]=1$; that is, both the columns can have non-zero entry simultaneously at maximum one position. Some authors refer to this condition as no more than one overlapping 1s.
\item[$(IV)$] Let $t$ be the weight of a column of $C$ and $t_{r}$ be a weight of a row of $C$ then $t \cdot t_{r} \leqslant p-1$. 
\end{description}

Now we prove bounds on $Aut(M)$ and minimal degree using a sequence of lemmas. First we just point out some relation between columns of matrices.

Let $P\in Aut(M)$ then for some $A$ we have $A \left[ I|C\right]P = \left[A|AC \right]P = \left[ I|C\right]$. Hence, $\left[ A|AC\right]$ have same set of columns as $\left[I|C \right]$ possibly in different order.
\begin{remark} \label{r5.1} Every column of $A$ and $AC$ is either a column of $C$ or a column of $I$. Also no column of $A$ is same as column of $AC$, in fact, no two columns of $\left[A|AC\right]$ are identical. \end{remark}
Assume for the whole discussion that every column of $C$ has weight $t$.

\begin{lemma}\label{l5.1} Let $\lbrace v_{1},v_{2},\ldots , v_{t}\rbrace$  be set of $t$ distinct columns where each $v_{i}$ comes either from $I$ or from $C$ such that at least 2 columns are from $C$ then $\Sigma_{i=1}^{t} v_{i}$ is of weight at least $2$.
\end{lemma} 
\begin{proof}
Suppose $v_{1},v_{2}$ are from $C$. Hence $v_{1}+v_{2}$ has weight at least $2t-2$ from condition (III) on $C$ (as at most one entry from each column can get converted to 0). Now each of the remaining $t-2$ columns $v_{3},v_{4},\ldots,v_{t}$ can reduce the weight by at most 2 as it can reduce weight of $v_{1}$ by at most 1 and weight of $v_{2}$ by at most 1. Thus weight of $\Sigma_{i=1}^{t} v_{i}$ is at least $2t-2-2(t-2)=2$.
\end{proof}
\begin{lemma} \label{l5.2} Let $\lbrace v_{1},v_{2},\ldots , v_{t}\rbrace$ be a set of $t$ distinct columns where each $v_{i}$ comes from either $I$ or $C$ such that $\Sigma_{i=1}^{t} v_{i}$ is weight 1. Then only possible combination is 1 column of weight $t$ and  $t-1$ columns of weight 1. Moreover, each column of weight 1, including the resultant $\Sigma_{i=1}^{t} v_{i}$, should have 1 in the same place as those of weight $t$ column.
\end{lemma}
\begin{proof}
Clearly, if all $v_{i}'s$ are weight 1, then $\Sigma_{i=1}^{t} v_{i}$ would be weight $t$. Now, if at least two $v_{i}'s$ are weight $t$ then $\Sigma_{i=1}^{t} v_{i}$ can not be weight 1 from $Lemma\  \ref{l5.1}$. The condition on position of 1 is easy to check so we skip the proof.
\end{proof}
\begin{remark} \label{r5.2} Since columns of $C$ have weight $t$, any column of $AC$ is an addition of $t$ columns of $A$. Moreover, every column of $A$ contributes to such $t_{r}$ additions where $t_{r}$ is the weight of a row of $C$.
\end{remark}
\begin{theorem}\label{t5.1} If $P\in Aut(M)$ with $C$ satisfying condition (II) and (III) then for any corresponding $A$, $AC$ can not have a column of weight 1.
\end{theorem}
\begin{proof}
Suppose there is a column (say $ac_{1}$) of $AC$ with weight 1. As columns of $AC$ are obtained by adding $t$ columns of $A$, there exists a set $ \lbrace a_{1},a_{2},\ldots,a_{t} \rbrace$  of $t$ distinct columns such that $\Sigma a_{i}=ac_{1}$.

From Lemma $\ref{l5.2}$, one column should be weight $t$ and rest of the columns must be weight $1$. Let $a_{1}$ be weight $t$ column and $a_{2},a_{3},\ldots,a_{t}$ be weight 1 and each weight 1 column matches with weight $t$ column.

Now since $a_{1}$ must be involved in $t_{r}$ such additions ($t_{r} \geq 2$), there exists another set $\lbrace a_{1},a'_{2},a'_{3},\ldots a'_{t} \rbrace$ such that $a_{1}+a'_{2}+a'_{3}+\cdots+a'_{t}=ac_{2}$ is another column of $AC$. As $ac_{2}$ is column of $AC$ it must be of weight 1 or weight $t$.

Observe that $ac_{2}$ can not be weight 1. Because if $ac_{2}$ was weight 1, then from Lemma $\ref{l5.2}$ it should match 1, with $a_{1}$. The only columns satisfying such condition are $a_{2},a_{3},\ldots,a_{t},ac_{1}$. But $ac_{2}$ can not be equal to either of those since column of $AC$ can not be equal to any column of $A$ or any other column of $AC$, this follows from Remark $\ref{r5.1}$.

Now only possibility is $ac_{2}$ has weight $t$. Now we analyze this possibility in two cases.

\textit{Case 1:} All $a'_{2},a'_{3},\ldots,a'_{t}$ have weight $t$. This leads to a contradiction as now the columns in the addition are weight $t$, all of them come from $C$. Thus we have one column of $C$ as sum of other columns of $C$. Contradiction to condition (II).

\textit{Case 2: } There is a column of weight 1 in addition (say $a'_{2}$). Now we have

$a_{1}+a'_{2}+a'_{3}+\cdots+a'{t}=ac_{2}$

Therefore, $a_{1}+ac_{2}+a'_{3}+a'_{4}+\cdots+a'_{t}=a'_{2}$. Contradiction to Lemma $\ref{l5.1}$ as $a_{1},ac_{2}$ have weight $t$ and $a'_{2}$ has weight 1.

Thus we have ruled out every possibility for $ac_{2}$. Hence our assumption that there  was a column of weight 1 in $AC$ is wrong. Hence, every column of $AC$ is weight $t$.
\end{proof}

\begin{corollary} \label{c5.1} If $C$ satisfies conditions (II),(II) and $M=\left[I|C \right]$ then every $P\in Aut(M)$ is a direct sum of permutation matrix of size $mp$ and a permutation matrix of size $p$ that is $P$ can be expressed as $P=P_{1} \oplus P_{2}$ where $P_{1}$ is a $mp \times mp$ permutation matrix and $P_{2}$ is a $p \times p$ permutation matrix.
\end{corollary}
\begin{proof}
Since $AC$ has no column of weight 1; $A$ must be permuted to get identity matrix and $AC$ must be permuted to get $C$. Thus $P$ permutes first $mp$ columns within themselves and last $k$ columns within themselves. Therefore, $P=P_{1} \oplus P_{2}$ with corresponding sizes. Moreover, $A={P_{1}}^{-1}$ and $ACP_{2}=C$.
\end{proof}
\begin{lemma} \label{l5.3} If $M$ satisfies condition (II) and (III) then $\vert Aut(M) \vert = \# (\mathcal{P}_{1},\mathcal{P}_{2})$ satisfying $\mathcal{P}_{1}C\mathcal{P}_{2}=C$.
\end{lemma} 
\begin{proof}
From corollary \ref{c5.1}; $A=P_{1}^{-1}$ and $ACP_{2}=C$. Therefore, $P_{1}^{-1}CP_{2}=C$ and satisfies required equation with $\mathcal{P}_{1}=P_{1}^{-1}$ and $\mathcal{P}_{2}=P_{2}$. Similarly, one can go back to show other side. Thus, $Aut(M)$ is in one-to-one correspondence with ($\mathcal{P}_{1}, \mathcal{P}_{2}$) solutions for $\mathcal{P}_{1}C\mathcal{P}_{2}=C$.
\end{proof}
\begin{corollary} \label{c5.2} If $M$ satisfies condition (II) and (III) then $\vert Aut(M) \vert = \# \mathcal{P}_{2}$ satisfying $\mathcal{P}_{1}C\mathcal{P}_{2}=C$.
\end{corollary}
\begin{proof}
Notice that since no two rows of $C$ are identical $C\mathcal{P}_{2}$ are identical as $\mathcal{P}_{2}$ just permutes columns. Hence there is at most one way to permute rows to get back $C$. Hence for every $P_{2}$ there is at most one $\mathcal{P}_{1}$. Since $\mathcal{P}_{2}$ satisfies the equation, we have a unique $P_{1}$ corresponding to $\mathcal{P}_{2}$.
\end{proof}

So due to corollary \ref{c5.2} problem of finding size of the automorphism group is reduced to finding number of $\mathcal{P}_{2}$ solutions to \begin{equation} \label{eq1}
\mathcal{P}_{1}C\mathcal{P}_{2}=C
\end{equation}
Here, we state a theorem by Burnside ~\cite[Theorem 3.5B]{Dixon}.
 \begin{theorem} \label{burni}
 \textsf{Burnside} Let $\mathrm{G}$ be a subgroup of $Sym(\mathbb{F}_{p})$ containing a $p-cycle$ $\mu : \xi \mapsto \xi+1$. Then $\mathrm{G}$ is either 2-transitive or $\mathrm{G} \leq AGL_{1}(\mathbb{F}_{p})$ where $AGL_{1}(\mathbb{F}_{p})$ is the affine group over $p$.
 \end{theorem}
We use this theorem to prove our main result
\begin{theorem}
\label{main_thm}
If $M$ satisfies conditions (I), (II) and (III) then $\vert Aut(M)\vert \leq p (p-1)$
\end{theorem}
\begin{proof}Let $\mathrm{G} \text{ be the set of } \mathcal{P}_{2}$ that satisfy equation \eqref{eq1}. It is an easy check that $\mathrm{G}$ forms a group. Also we can check that $\mathcal{P}_{2}=\mu \in \mathrm{G}$ as it satisfies the equation \eqref{eq1} with $\mathcal{P}_{1}$ being block diagonal with $\mu^{-1}$ as every block. So using Burnside's theorem we can say that $\mathrm{G}$ is a subgroup of $ AGL_{1}(\mathbb{F}_{p})$ if it is not doubly transitive and the its size is less that or equal to $p(p-1)$. Proof that $\mathrm{G}$ is not doubly transitive follows from condition $(IV)$. 
\end{proof} 
\begin{lemma} If $t \cdot t_{r} \leqslant p-1$ then $\mathrm{G}$ is not doubly transitive.
\end{lemma}
\begin{proof}
Let $\mathcal{S}$ be the set of $x$ such that there exists a row $r$ in $C$ which contains non-zero entries at positions $0$ and $x$ that is there exists a row $r$ such that $r[0]=r[x]=1$. Now clearly $\vert \mathcal{S}\vert \leqslant t \cdot t_{r}$. Hence, $\vert \mathcal{S}\vert < p$ and there exists $y\in \lbrace 0,1,2,\ldots, p-1 \rbrace$ such that $y\notin \mathcal{S}$. Denote first row of $C$ by $r_{0}$. Let $x_{0},y_{0}$ be distinct positions such that $r_{0}[x_{0}] = r_{0} [y_{0}] = 1$. Now one can notice that no element $\mathcal{P}_{2} \in \mathrm{G}$ can send $(x_{0},y_{0})$ to $(0,y)$ as $C\mathcal{P}_{2}$ needs to have same set of rows as $C$.  After action of such  $\mathcal{P}_{2}$ first row has 1's at positions $0$ and $y$ but no row can have 1's at positions $0$ and $y$ the reason being $y \notin \mathcal{S}$. Thus $\mathrm{G}$ is not 2 transitive.
\end{proof}
\begin{lemma} Minimal degree of $\mathrm{G}$ is at least $p-1$.
\end{lemma}
\begin{proof}
As $ P = P_{1} \oplus P_{2}$ number of points moved by $P$ is greater than or equal to number of points moved by $P_{2}$. Now assume $P_{2}$ fixes at least two points. Then $P_{2} = I_{p}$ as $P_{2} \in AGL(\mathbb{F}_{p})$. And as for every $P_{2}$ there is at most one corresponding $P_{1}$, $P_{1} = I_{(m-1)p}$ making $P= I_{mp}$. Thus non identity $P$ can not have corresponding $P_{2}$ fixing more than one point and hence every non identity $P$ moves at least $p-1$ points. 
\end{proof}
\begin{algorithm}
\caption{An algorithm that generates required generator matrix}
\begin{algorithmic}[1]
\Procedure{$Generate\_C$}{$p,m,t_{r}$} 
\State $Available\_set=\left[ 0,1,\ldots, p-1 \right]$
\State $\mathcal{A}= \left[ \; \right]$
\State \textbf{repeat $m$ times} 
$\Big{\lbrace}$ \While{size of Current\_set $\leq t_{r}$}
\State randomly choose $\alpha_{1} \in$ Available\_set
\State Current\_set.append[$\alpha_{1}$]
\State remove $\alpha_{1}$ from Available\_Set
\For { $\left( \alpha_{2}, \alpha_{3}\right) \in Current\_set \cup \mathcal{A}$ }
\State remove $\alpha_{2}+\alpha_{3}-\alpha_{1}\; modulo\; p$ from Available\_set
\EndFor
\EndWhile
\State $\mathcal{A}.append\left(Current\_State \right) \Big{\rbrace}$
\State construct circulant $C_{i}$ having $\mathcal{A}[i]$ as its first column
\State \textbf{return} $C = $ \textsc{Stack} $\left[ C_{0},C_{1},\ldots,C_{m-1}\right]$
\EndProcedure
\end{algorithmic}
\end{algorithm}

We will now prove correctness of the algorithm. We claim that output of the algorithm is a $C$ satisfying condition $(III)$. Let $\mathcal{A} = \lbrace \alpha_{j}\rbrace$ denote positions such that $c_{0} [\alpha_{j}]=1$ where $c_{0}$ is the first column of $C$.
\begin{lemma} Let $\mathcal{A}^{\prime}= \lbrace \alpha^{\prime}_{j}\rbrace$ denote positions such that $c_{N}[\alpha^{\prime}_{j}] = 1 $ where $c_{N}$ denotes the $N+1\,th$ column of $C$ then $\lbrace  \alpha_{j} + N\; mod\; p\; |\; \alpha_{j} \in \mathcal{A} \rbrace = \lbrace \alpha^{\prime}_{j} \;|\; \alpha^{\prime}_{j} \in  \mathcal{A}^{\prime} \rbrace$.
\end{lemma}  
\begin{lemma} If condition $(III)$ is not satisfied by $C$ then there exists $\alpha_{1}, \alpha_{2}, \alpha_{3}, \alpha_{4}$ such that $\alpha_{4} = \alpha_{1}+\alpha_{3}- \alpha_{2} \; mod\; p$ such that $\alpha_{1}, \alpha_{2}$ belong to the same circulant $C_{i}$ and $\alpha_{3}, \alpha_{4}$ belong to the same circulant $C_{j}$ \footnote{$C_{i}$ can be same as $C_{j}$ that is it is possible that $i = j$ where the case represents both the overlaps happening in the same circulant}.
\end{lemma} 
\begin{proof}
Without loss of generality we can assume that column $c_{0}$ and column $c_{N}$ have more than one overlaps. So there exists $\alpha_{1}$ such that $c_{0}\left[ \alpha_{1} \right] = c_{N} \left[ \alpha_{1} \right] = 1$. Now from second equality and above lemma $\alpha_{1} = \alpha_{2} + N \; mod \; p$ for some $\alpha_{2} \in \mathcal{A}$ . Similarly second overlap will give you $\alpha_{3} = \alpha_{4} + N\; mod \; p$. Solving both the equations we get $\alpha_{4} = \alpha_{1}+\alpha_{3}- \alpha_{2} \; mod \; p$.
\end{proof}
Our algorithm iteratively constructs circulants $C_{0},C_{1},\ldots,C_{m-1}$ so that $\alpha_{4} \neq \alpha_{1}+\alpha_{3}- \alpha_{2}$. In this way we ensure that $C$ generated as an output of algorithm satisfies condition $(III)$.

We end this chapter by showing how to achieve other conditions. Condition $(I)$ is trivial as we just need to choose a prime $p$. We now move towards condition $(II)$. By Lemma $\ref{cir2}$ we know that $\mathcal{C}_{p}$ is isomorphic to direct product of two rings. First part $\frac{\mathbb{F}_{q}\left[ x
\right]}{(x-1)}$ is a field as $(x-1)$ is irreducible. So to ensure that image in this ring is non-zero we just have to ensure that weight of circulant is odd. So we get our first condition as $t_{r}$ should be odd. For the second part of the product we use following lemma.
\begin{lemma} If $2$ is a primitive root modulo $p$ then $\Phi_{p}(x)$ is irreducible in $\mathbb{F}_{2} \left[x \right]$.
\end{lemma}
Proof of this can be found in standard number theory text.
So to make the second part a field we need to choose $p$ such that $2$ is primitive. One of the ways to do this is choose $p=4q+1$ such that both $p$ and $q$ are prime. For condition $(IV)$ notice that $t = m\cdot t_{r}$ which gives us $t \cdot t_{r} = {t_{r}}^2 \cdot m$. So we can satisfy condition $(IV)$ by choosing $t_{r} \leqslant \sqrt{\frac{p}{m}}$.
 
In conclusion, we can generate a matrix $C$ satisfying required conditions by running algorithm for prime $p$ such that $2$ is primitive mod $p$ and choosing odd $t_{r}$ less than $\sqrt{\frac{p}{m}}$ and thus we can construct $\frac{m-1}{m}$ codes with small automorphism group and minimal degree at least $p-1$.

%% file: Chapter4.tex
This chapter is based on work done in ~\cite{Our}
\section{Introduction}
In chapter 2 we mentioned a way to ensure security of cryptosystem against quantum Fourier sampling. But before going there, we describe our variant and briefly go over its classical security. Our variant is based on Quasi-cyclic codes over $\mathbb{F}_{2^{l}}$ of rate $\frac{m-1}{m}$. The relevant definitions about Quasi-cyclic codes are in section 3.1 and more details are in ~\cite{Gulliver_thesis}. 

\subsection{Description of our Niederreiter cryptosystem} \paragraph{Description of the parity check matrix used for the proposed Niederreiter cryptosystem}\label{dd}
Recall that we are talking about $\frac{m-1}{m}$ quasi-cyclic codes over $\mathbb{F}_{2^l}$. For the cryptosystem to be quantum-secure the parity check matrix $\mathcal{H}$ for the $[n=mp,k=(m-1)p,d]$, $\frac{m-1}{m}$ quasi-cyclic code should satisfy the following conditions:
\begin{itemize}
\item[I] Integers $m,p$, such that $p$ is a prime and $m$ is bounded above by a polynomial in $p$.
\item[II] The matrix $\mathcal{H}$ is of size $p\times mp$ over $\mathbb{F}_{2^l}$.
\item[III] The matrix $\mathcal{H}$ is of the form $\left[\,C_0=I\,|\,C_1\,|\,C_2\,|\ldots\,|\,C_{m-1}\,\right]$, where each $C_i$ is a circulant matrix of size $p$. Each $C_i$ for $i>0$ should contain an element from a proper extension of $\mathbb{F}_2$.
Furthermore, we denote the matrix $\mathcal{H}$ as $\left[\,I\,|\,C\,\right]$ where $C$ is the concatenation of the circulant matrices $C_i$, $i>0$.
\item[IV] We define $T_{\mathcal{H}}=\left\{P_1\in \mathrm{S}_p \;|\; \exists P_2\in \mathrm{S}_{p(m-1)} \;\text{such that}\; P_1CP_2=C\right\}$, where $\mathrm{S}_n$ is the symmetric group acting on $n$ letters. It is easy to see that $T_\mathcal{H}$ is a permutation group action on $p$ letters. The condition we impose on $\mathcal{H}$ is that $T_\mathcal{H}$ is not 2-transitive.
\item[V] No two columns of $C$ are identical. 
\end{itemize}
\section{Classical Attacks}
In this section we briefly go over the generic classical attacks against McEliece and Niederreiter cryptosystems. We also mention some attacks exploiting the circulant structures in the keys. Interestingly, Li \textit{\etal}~\cite[Section III]{Li} proved that both McEliece and Niederreiter cryptosystems are equivalent in terms of classical security. The proof follows from the fact that the encryption equation for one can be reduced to the other. This implies the equivalence of security of both the cryptosystems for attacks that try to extract the plaintext from a ciphertext.
  
Most generic attacks over algebraic code based cryptosystems are \emph{information set decoding attacks}(ISD). Two most popular ways of implementing ISD attacks are by Lee and Brickell~\cite{Lee} and Stern~\cite{Stern}. As mentioned by Baldi \etal~\cite{Baldi} ISD attacks are the best known attacks with the least work factor as far as classical cryptanalysis is considered. Hence these work factors are considered as security levels for a McEliece and Niederreiter cryptosystems.  

The basic idea behind one of the attacks was suggested by McEliece himself. Lee and Brickell~\cite{Lee} improved the attack and added an important verification step where attacker confirms that recovered message is the correct one. In this case, we are dealing with a McEliece cryptosystem over a $[n,k]$ linear code. The strategy is based on repeatedly selecting $k$ bits at random from a $n$-bit ciphertext in hope that none of the selected bits are part of the error. Similar attacks can also be implemented over Niederreiter cryptosystems.  Lee and Brickell also provided a closed-form equation for complexity of the attack. As our system is based on $\left( n=mp, k=(m-1)p, d_{min} = 2\mathcal{E}+1\right)$ code the expression for minimal work factor (with $\alpha = \beta = 1$ as taken by Lee and Brickell) takes the following form $$
W_{min}= W_{2}= T_{2}  \left( (m-1)^{3} p  ^{3} + (m-1) p N_{2} \right)$$ where $T_{2}= \dfrac{1}{Q_{0} + Q_{1} + Q_{2}}$ and $Q_{i}=$ $\mathcal{E} \choose i$ ${n-\mathcal{E}} \choose {k-i} $ / $n \choose k$ with $N_{2}= 1 + k + {k \choose 2}$.

In Table~\ref{table} we present numerical data for work factor for different values of parameters. Recently, Aylaj \textit{\etal}~\cite{Aylaj} developed an algorithm to construct stack-circulant codes with high error correcting capacity which makes the proposed Niederreiter cryptosystem much more promising.

Other ISD attacks are based on a strategy given by Stern. To recover the intentional error vector $e$ in a McEliece cryptosystem such strategies use an extension code $C^{\prime \prime}$ generated by generator matrix $M^{\prime \prime}=\left[ \begin{array}{c}
M^\prime \\
x 
\end{array} \right]$.  Bernstein \etal~\cite{Bernstein} later improved this attack. Probability of success and work factor for Stern's attack is described in ~\cite{Hirotomo}. In the Table~\ref{table} we also provide probability of success for parameters $l=16$ and $A_{w} \approx n-k$. Both the parameters can be optimized further to obtain the least work factor but not much variation is seen as we change any of these parameters. With such low probabilities, it is clear that the work factor for Stern's attack is worse than the Lee-Brickell attack. Even when one considers improvements suggested by Bernstein \textit{\etal}~\cite{Bernstein}, Lee-Brickell's~\cite{Lee} attack seems to outperform the attack by Bernstein \textit{\etal}~as it produces speedup up to 12 times and hence the security of the system against the Lee-Brickell attack should be considered the security of the system. Key sizes should be devised according to that. 

Another attack worth mentioning for quasi-cyclic codes is the attack on the dual code. This attack works only if the dual code has really low weight codewords and is often encountered only when sparse parity check matrices are involved. For example, McEliece with QC-LDPC~\cite{Baldi}. Such attacks can easily be stopped by choosing codes that do not have low weight codewords. From the work of Aylaj \etal~\cite{Aylaj} this can be achieved.

\section{Quantum security}
After this discussion on classical security we now move towards quantum-security of the proposed McEliece and Niederreiter cryptosystems which is one of the major goal of this chapter.
\
Before moving towards quantum security we recall definitions and theorem by Dinh, Moore and Russell. 

\begin{definition} $Aut(M)=\lbrace P\in S_{n}$ such that  there exists $ A \in GL_{k}(\mathbb{F}_{q}) ,\ AMP=M \rbrace $
 \end{definition}
\begin{definition} The minimal degree of a $G \leqslant S_{n}$ acting on set of $n$ symbols is defined to be minimum number of elements moved by a non-identity element of the group $G$.
\end{definition}
\begin{definition}Consider a  $k\times n$ matrix $M$ , we define $T_M$ for matrix $M= \left[I_{k} | M^{*} \right]$ as 

$T_M=\lbrace \mathcal{P}_{1} \in S_{k}$ such that there exists $\mathcal{P}_{2}\in S_{n-k}$ with $\mathcal{P}_{1} M^{*} \mathcal{P}_{2} =M^{*}$ $\rbrace$
\end{definition}
\begin{theorem} \label{thm1}
~\cite[Theorem 4]{Dinh}: Assume $q^{k^2} \leqslant n^{an}$ for some constant $0 < a < 1/4$. Let $m$ be the minimal degree of the automorphism group $Aut(M)$. Then for sufficiently large n, the subgroup $\mathrm{K}$, $D_{K} \leqslant O(\vert K \vert ^ 2 e^{-\delta m} ) $, where $\delta > 0$ is a constant.
\end{theorem}

The idea behind security of the system is when distinguishability of the hidden subgroup $\mathrm{K}$ denoted as $D_{k}$ becomes less than $\log^{-\omega(1)} \vert \mathrm{G}\vert$, quantum Fourier sampling can not give us the hidden subgroup and hence an attacker can not find a scrambler permutation pair.
\subsection{Proof of indistinguishability of the hidden subgroup}
We prove indistinguishability in a sequence of lemmas.
\begin{lemma} \label{ov1} Let $P \in \rm{Aut}(\mathcal{H})$ then $P= P_{1} \oplus P_{2}$ where $P_{1}$ is a block of size $p$ and $P_{2}$ is a block of size $(m-1)p$ and $P_1\oplus P_2$ is a block diagonal matrix of size $mp\times mp$ with the top block $P_1$ and the bottom block $P_2$.
\end{lemma}
\begin{proof} Let $P \in \rm{Aut}(\mathcal{H})$, from the definition of automorphism there is an $A$ such that $A\mathcal{H}P=\mathcal{H}$. 
This implies that
\[A \left[\,I\,|\,C\,\right] P = \left[\,A\,|\,AC \,\right]P= \left[\,I\,|\,C\, \right].\] 

As action of right multiplication by a permutation matrix permute columns, the above equality shows that $\left[\,A\,|\,AC\,\right]$ has same columns as $\left[\,I\,|\,C \,\right]$ possibly in different order. Now since every column of $C$ contains an entry from a proper extension of $\mathbb{F}_q$, no column of $A$ can be column of $C$. This forces $A$ to have same columns as $I$ and $AC$ to have same columns as that of $C$. Hence $P$ permutes first $p$ columns within themselves and last $(m-1)p$ columns in themselves. Hence every $P \in \rm{Aut}(\mathcal{H})$ can be broken into $P_{1} \oplus P_{2}$ so that $P_{1}$ acts on $I$ and $P_{2}$ acts on $C$. 
\end{proof} 
\noindent An obvious corollary follows:
\begin{corollary}\label{ov2} If $P \in \rm{Aut}(\mathcal{H})$ then the corresponding $A \in \mathrm{S}_{k}$.
\end{corollary}
\noindent The next lemma is central to quantum-security. It gives us a way to move from $\mathcal{H}$ to $C$ by noting, the $P_1$ from the $P\in\rm{Aut}(\mathcal{H})$ is actually a member of $T_{\mathcal{H}}$.
\begin{lemma} \label{ov3} The cardinality of $\rm{Aut}(\mathcal{H})$ is the cardinality of the set $\left\{(P_{1},P_{2})\right\}$ that satisfy $P_{1}C P_{2}=C$ where $\mathcal{H}=\left[\,I\,|\,C\,\right]$ as defined earlier.
\end{lemma}
\begin{proof} 
The proof follows from the fact, if $P$ belongs to $\rm{Aut}(\mathcal{H})$, then $P=P_1\oplus P_2$. Then $A\left[\,I\,|\,C\,\right]P=\left[\,I\,|\,C\,\right]$ translates into $A\left[\,I\,|\,C\,\right](P_1\oplus P_2)=\left[\,I\,|\,C\,\right]$.  Keeping in mind the block diagonal nature of $P$, it follows that $\left[\,AIP_1\,|\,ACP_2\,\right]=\left[\,I\,|\,C\,\right]$. Then $A=P_1^{-1}$ and $P_1^{-1}CP_2=C$. This proves the lemma.
\end{proof}
\noindent The next lemma proves that for each $P_1$ there is at most one $P_2$.
\begin{lemma}\label{ov4} Cardinality of the set $\lbrace\left(P_{1},P_{2} \right)$ that satisfy $P_{1}C {P}_{2}=C\rbrace$ equals $\vert T_{\mathcal{H}} \vert$.  
\end{lemma}
\begin{proof} Recall that $T_{\mathcal{H}} =  \lbrace P_{1}$ that satisfy $P_{1}CP_{2}=C \rbrace$. So it suffices to show that for every $P_{1}$ there is at most one $P_{2}$. Since no two columns of $C$ are identical, no two columns of $P_{1}C$ are identical. Hence, there is at most one way to re-order them to get back $C$. Thus for every $P_{1}$ there is at most one $P_{2}$.
\end{proof}
\begin{theorem}[Burnside~{\cite[Theorem 3.5B]{Dixon}}] \label{burni}
Let $G$ be a subgroup of $Sym(\mathbb{F}_{p})$ containing a $p$-cycle $\mu : \xi \mapsto \xi+1$. Then $G$ is either 2-transitive or $G \leq A\rm{\rm{GL}}_{1}(\mathbb{F}_{p})$ where $A\rm{\rm{GL}}_{1}(\mathbb{F}_{p})$ is the affine group over $p$.
 \end{theorem}
\noindent We prove a theorem on the size of the \rm{Aut}omorphism group of $\mathcal{H}$. 
\begin{theorem}
If $\mathcal{H}$ satisfies conditions I,II and III then $\vert \rm{Aut}(\mathcal{H}) \vert \leqslant p(p-1)$.
\end{theorem}
\begin{proof}
From Lemma~\ref{ov3} and Lemma~\ref{ov4}, the group $\rm{Aut}(\mathcal{H})$ has same size as $T_{\mathcal{H}}$. It is now easy to check that the circulant matrix $\mu$ with first row $[0,1,0,\ldots,0]$ of size $p$ belongs to $T_{\mathcal{H}}$. The corresponding $P_{2}$ will be a block diagonal $(m-1)p$ matrix with blocks of size $p$ and each consisting of $\mu^{-1}$. Now notice that the circulant matrix $\mu$ corresponds to the $p$-cycle $\xi\mapsto\xi+1$.
By our condition III, $T_{\mathcal{H}}$ is not 2-transitive. Now by Burnside's theorem $T_{\mathcal{H}} \leq A\rm{\rm{GL}}_{1}(\mathbb{F}_{p})$. Thus $\vert \rm{Aut}(\mathcal{H}) \vert \leqslant p (p-1)$.
\end{proof}
\noindent After this bound on the size of the \rm{Aut}omorphism group we move towards the minimal degree of the \rm{Aut}omorphism group.

\begin{lemma} The minimal degree of $\rm{Aut}(\mathcal{H})$ is bounded below by $p-1$.
\end{lemma}
\begin{proof}
Notice that any $P\in\rm{Aut}({\mathcal{H}})=P_1\oplus P_2$. By the twist, from $P\in\rm{Aut}(\mathcal{H})$ to $P_1^{-1}\in T_{\mathcal{H}}$, it is easy to see that $P_{1} \in A\rm{GL}_k(\mathbb{F}_{q})$. Then $P_{1}(x)=ax+b \pmod{q}$ for some $a,b\in\mathbb{F}_q$. If $P$ fixes two distinct points, then $a=1$ and $b=0$ is the only possible solution. This corresponds to the identity element and thus a non-identity element can not fix more that one point. So minimal degree of  $\rm{Aut}(\mathcal{H})$ is bounded below by $p-1$.
\end{proof}

\noindent We now prove the main theorem of this chapter.
\begin{theorem}\label{mainthm}
Let $p$ be a prime and $m$ a positive integer bounded above by a polynomial in $p$, such that, $p\leq\frac{1}{4}m\left(\log{m}+\log{p}\right)$. Then the subgroup $K$ (Equation~\ref{eqnK}) defined above is indistinguishable.
\end{theorem}
\begin{proof}
We will use Theorem~\ref{thm1} in this proof. First note, the minimal degree is bounded below by   
$p-1$. Now it is well known that $|K|=2|H_0|^2$ and $|H_0|=|\rm{Aut}(\mathcal{H})|\times|\rm{Fix(\mathcal{H})}|$. We have shown that $|\rm{Aut}(\mathcal{H})|\leq p(p-1)$ and it is easy to see that $|\rm{Fix(\mathcal{H}})|=1$. Putting all these together, we see that $|K|^2e^{-\delta p}\leq 4p^8e^{-\delta p}$ for some positive constant $\delta$. However, from the bound on the size of $m$, it is obviously true that $4p^8e^{-\delta p}\leq \left(mp\log{(mp)}\right)^{-\omega(1)}$ for large enough $p$.

Now, if $p\leq am\left(\log{m}+\log{p}\right)$, then $p^2\leq amp\left(\log{m}+\log{p}\right)$ which gives $2^{p^2}\leq(mp)^{amp}$ for $0<a<\frac{1}{4}$. This satisfies the premise of Theorem~\ref{thm1} and hence $K$ is indistinguishable.
\end{proof}

\section{Construction of the required parity check matrix}
Now we address the last question about the proposed \N cryptosystem, how to construct a matrix $\mathcal{H}$ satisfying conditions I, II, III and IV? Clearly, conditions I, II and III are trivial to set up and deserve no special attention. We suggest a particular way for construction of parity check matrix $\mathcal{H}$ so that condition IV is satisfied. It should be noted that there may be other ways to meet condition IV as well. 

Choose a pair of distinct elements $a,b\in F_{q^l}$. Now construct $\mathcal{H}$ such that $C_{1}$ contains both $a$ and $b$ exactly once in each column and no other $C_{i}$ contains both $a$ and $b$. We restate this condition as our condition $\mathrm{IV}^\prime$. We could have replaced $C_{1}$ by any other $C_{i}$ for $i>1$ and the proof remains the same. For sake of simplicity we stick with $C_{1}$.
\begin{description}
\item[$\mathrm{IV}^\prime$] Two distinct elements $a,b\in F_{q^l}$ occurs as entries of $C_{1}$ exactly once in each column and no other $C_i$ contain both $a$ and $b$.
\end{description}

\begin{lemma} If the matrix $\mathcal{H}$ satisfies $\mathrm{IV}^\prime$, it also satisfies $\mathrm{IV}$.
\end{lemma}
\begin{proof} Let $\mathcal{P}_{1} \in T_{\mathcal{H}}$. From $\mathcal{P}_1 C\mathcal{P}_2=C$ it follows
 that $\mathcal{P}_{1}C$ should have the same set of columns as $C$ but possibly in a different order. Let
  $\alpha$ denote the row of $a$ in the first column of $C_{1}$ and $\beta$ denote the row of $b$ in the
   same column. Now notice that every column in $C$ that contains both $a$ and $b$ contains them such that
    difference between rows of $a$ and $b$  is $\alpha - \beta$ mod $p$ where $p$ is the size of each
     circulant matrix. Now let $\sigma\in T_{\mathcal{H}}$ such that it sends $\beta$ to $\alpha$ and 
     $\alpha$ to $\beta$. It then follows from the fact that $p$ is a odd prime, $\alpha=\beta$ which
      contradicts our assumption. Hence, $T_{\mathcal{H}}$ is not 2 transitive. 
\end{proof}
Condition V can be easily satisfied using brute force and other means and this completes the construction of a parity check matrix $\mathcal{H}$ satisfying I, II, III, IV and V and hence, a \textbf{Niederreiter cryptosystem that resists quantum Fourier sampling} is found.
\
\begin{algorithm}
\caption{An algorithm that generates required parity check matrix}
\begin{algorithmic}[1]
\Procedure{$Generate\_H$}{$p,m$}
\State choose $a,b \in \mathbb{F}_{2^{l}}^{\times}$
\State generate an array $v$ of length $p-2$ with entries from $\mathbb{F}_{2^{l}}- \lbrace a,b \rbrace$
\State Construct an array $c_{1}$ by concatenating array $[a,b]$ with the array $v$ 
\State randomly permute entries within $c_{1}$
\State Let $I$ denote the array of length $p$; $I = \left[1, 0, 0,\ldots, 0 \right]$
\State $vector\ list = \left[ I, c_{0}\right]$
\State Let $R$ denote the right shift operator
\State construct an array $x$ of length $p$ over $\mathbb{F}_{{2}^{l}}$
\While{$vector\ list$ has length less than $m$}
\For{$y \in vector\ list$}
\For{$0\leqslant i \leqslant n-1$}
\If {$R^{i}(x) = y$} 
\State choose next $x$
\Else
\State add $x$ in the $vector\ list$
\State choose next $x$
\EndIf
\EndFor
\EndFor
\EndWhile
\For{$v_{i} \in vector\ list\ ; 0\leqslant i \leqslant m-1$}
\State Construct a circulant matrix $C_{i}$ with $v$ as its first column
\EndFor
\State \textbf{return} parity check matrix $H=$ \textsc{Array}$\left[ C_{0},C_{1},\ldots,C_{m-1}\right]$
\EndProcedure
\end{algorithmic}
\end{algorithm}

\section{Advantages of the proposed cryptosystem}
One of the prime advantages of our proposed cryptosystem is quantum-security. Apart from that it has high transmission rate which translated into high encryption rate. Current McEliece cryptosystem built on Goppa codes has transmission rate of about $0.52$. For a McEliece cryptosystem its rate is same as that of the transmission rate of the underlying code and is $\frac{k}{n}$. Niederreiter cryptosystems have slightly different rates due to difference in their encryption algorithm. For a general cryptosystem its encryption rate or information rate can be defined as follows ~\cite{NiedBook}:

Let $\mathcal{S}(C)$ denote possible number of plaintext and $\mathcal{T}(C)$ denote possible number of ciphertexts then information rate of the system is defined by 
\[\mathcal{R}(C) = \dfrac{log\left(\mathcal{S}(C)\right)}{log\left(\mathcal{T}(C)\right)}.\]
This information rate can be viewed as amount of information contained in one bit of ciphertext.  

Our proposed Niederreiter cryptosystem have encryption rate on the higher side. This gives our variant an edge over once those constructed on classical Goppa codes or with GRS codes (generalized Reed-Solomon codes).

As discussed before another problem with McEliece and Niederreiter cryptosystems is large key size. Circulant matrices seems like a good choice when it comes to key-sizes. Matrices are 2-dimensional objects but circulant matrices behave like a 1-dimensional object as they can be described by their first row. Though this circulant structure is lost in public key due to the scrambler-permutation pair, the size of the key still remains smaller than the conventional Niederreiter cryptosystem. Our system is slightly better than original Niederreiter cryptosystem because of the less number of rows in the public key matrices. With $p=101$, this number is less than one-tenth of the original Niederreiter cryptosystem. Though there are two factors that increase size of matrices in our variant compared to original McEliece, one, our matrices have large number of columns; and two, our system is based on extension field $\mathbb{F}_{q^{l}}$ which makes the effective size of the matrix $l$ times compared to McEliece which is based on $\mathbb{F}_{2}$. However, in most cases due to very less number of rows the net result indicates that our system requires shorter keys than original McEliece. For instance, at 80-bit security with $p=101$ and $l=3$ our keys are almost half of the keys corresponding to original McEliece at same security level. While at 256-bit security level with $p=211, t=40$ and $l=3$ our system key size of about $\frac{1}{4}^{th}$ of the original McEliece.  
\begin{table}[ht]
\caption{Parameters for the proposed \N cryptosystem}
\centering
\label{table}
\begin{tabular}{|l|c|c|c|c|c|c|c|c|c|c|}
\hline\hline
Security & $p$ & $t$ & $m_c$ & $m_Q$ & $m$ & Probability & \multicolumn{2}{|l|}{Public Key Size}& Rate\\
\cline{8-9}
in bits & & & & & & of success & No.~rows & No.~cols & \\
\hline
\multirow{4}{*}{80-bits}                                                         & \multirow{2}{*}{101} & 15                        & 17 & 35  & 35 & $2^{-132}$                                                           & 101                                         & 3535 & 0.60 \Tstrut\\
                                                                                 &                      & 20                        & 9              & 35 & 35& $2^{-190}$                                                          & 101                                         & 3535 & 0.77 \Tstrut\\
                                                                                 & \multirow{2}{*}{211} & 35                        & 4 & 62 & 62 & $2^{-398}$                                                          & 211                                         & 13082 & 0.71 \Tstrut\\
                                                                                 &                      & 40                       & 3& 62 & 62 & $2^ {-465}$                                                          & 211                                         & 13082  & 0.80                                     \TBstrut\\
                                                                                 \hline
\multirow{4}{*}{100-bits}                                                       & \multirow{2}{*}{101} & 15                        & 40      & 35     & 40  & $2^ {-136}$                                                          & 101                                         & 4040 & 0.61\Tstrut\\
                                                                                 &                      & 20                        & 17 &  35 & 35&$2^{-190}$                                                          & 101                                         & 3535 & 0.77\Tstrut\\
                                                                                 & \multirow{2}{*}{211} & 35                       & 5 &  62 & 62 & $2^{-398}$                                                          & 211                                         & 13082 & 0.71 \Tstrut\\
                                                                                 &                      & 40                       & 5 & 62 & 62 & $2^{-465}$                                                          & 211                                         & 13082 & 0.80 \TBstrut\\
                                                                                 \hline
\multirow{4}{*}{120-bits}                                                      & \multirow{2}{*}{101} & 15                        & 95 & 35   & 95         & $2^{-171}$                                                          & 101                                         & 9595 & 0.67\Tstrut\\
                                                                                 &                      & 20                        & 32 & 35 & 35& $2^{-190}$                                                          & 101                                         & 3535 & 0.77 \Tstrut\\
                                                                                 & \multirow{2}{*}{211} & 35                        & 8& 62 & 62 & $2^{-398}$                                                          & 211                                         & 13082 & 0.71 \Tstrut\\
                                                                                 &                      & 40                       & 6& 62 & 62 & $2^{-465}$                                                          & 211                                         & 13082 & 0.80 \TBstrut\\
                                                                                 \hline                                                                                                  
                                                                                 \multirow{4}{*}{128-bits}                                                      & \multirow{2}{*}{101} & 15                        & 134 & 35   & 134         & $2^{-184}$                                                          & 101                                         & 13534 & 0.70\Tstrut\\
                                                                                 &                      & 20                        & 42 & 35 & 42& $2^{-199}$                                                          & 101                                         & 4242 & 0.79 \Tstrut\\
                                                                                 & \multirow{2}{*}{211} & 35                        & 9& 62 & 62 & $2^{-398}$                                                          & 211                                         & 13082 & 0.71 \Tstrut\\
                                                                                 &                      & 40                       & 7& 62 & 62 & $2^{-465}$                                                          & 211                                         & 13082 & 0.80 \TBstrut\\
                                                                                 \hline
\multirow{2}{*}{256-bits}                                                      & \multirow{2}{*}{211} & 35                        & 98 & 62   & 98         & $2^{-443}$                                                          & 211                                        & 20678 & 0.75\Tstrut\\
                                                                                 &                      & 20                        & 55 & 62 & 62& $2^{-465}$                                                          & 211                                        & 13082 & 0.80 \Tstrut\\ \hline                                                                                                                                                                                             
\end{tabular}
\end{table}

%% file: Chapter5.tex
This thesis looks into two different variants of McEliece cryptosystem and corresponding NP-hard scrambler permutation problem from the point of view of security against quantum computers. In chapter 3 we attempted a problem of combinatorial optimization where the underlying code needs to have small automorphism group and large minimal degree so that we get $\vert Aut(M) \vert ^{4} e^{-\delta (m)} \leqslant log^{-\omega(1)}\vert \mathrm{G} \vert$.  This does not produce a McEliece cryptosystem as the premise of the theorem for quantum security requires $q^{k^{2}} \leqslant n^{\alpha n}$ for some $0 < \alpha < \frac{1}{4}$, the problem still is mathematically strong enough to deserve some attention. We provide a cryptosystem satisfying suggested bounds on automorphism group and minimal degree along with the algorithm to construct parity check matrices for this variant. It would be interesting to see if one can relax the requirement $q^{k^{2}} \leqslant n^{\alpha n}$ so that we get a quantum secure  McEliece over quasi-cyclic codes. Another direction of improvement could be refining condition for the construction. In particular, if condition $(III)$ can be relaxed it can increase the class of codes significantly.

In chapter 4 we give a Niederreiter variant that is classically and quantum secure against current known attacks. In particular, we show that for our system the hidden subgroup coming from natural reduction of corresponding scrambler-permutation problem is indistinguishable by quantum Fourier sampling. We also show that our system has higher encryption rate and shorter keys compared to classical McEliece systems. One of the important problem that needs to be addressed is finding QCCs that fit the suggested parameter size. It would be interesting to see if the system remains classically secure if we use  sparse keys. It is clear that the system remains secure against quantum computers as the group structure for the system remains the same. This is important because it could reduce key sizes substantially. Other than classical security, the real question is, can we construct codes using sparse parity check matrices simultaneously satisfying required conditions $(I)-(V)$ and retain error correction capacity of the system as par suggested parameters? Another problem of interest could be optimizing values of $p,t,l$ so that we get high rates and lower sized keys.

%% file: Thesis_Upendra.bbl
\begin{thebibliography}{10}

\bibitem{McEliece}
R.~J. McElice, ``A public key cryptosystem based on algebraic coding theory,''
  tech. rep., Communications system research centre, NASA, Jan-Feb 1978.

\bibitem{Blahut}
R.~E. Blahut, {\em Algebraic codes for data transmission}.
\newblock Cambridge University Press, 2003.

\bibitem{Roth}
R.~Roth, {\em Introduction to Coding Theory}.
\newblock New York, NY, USA: Cambridge University Press, 2006.

\bibitem{Huffman-Pless}
W.~C. Huffman and R.~A. Brualdi, {\em Handbook of Coding Theory}.
\newblock New York, NY, USA: Elsevier Science Inc., 1998.

\bibitem{brl}
E.~Berlekamp, R.~McEliece, and H.~Van~Tilborg, ``On the inherent intractability
  of certain coding problems (corresp.),'' {\em IEEE Transactions on
  Information Theory}, vol.~24, no.~3, pp.~384--386, 1978.

\bibitem{Petrank}
E.~Petrank and R.~M. Roth, ``Is code equivalence easy to decide?,'' {\em IEEE
  Transactions on Information Theory}, vol.~43, no.~5, pp.~1602--1604, 1997.

\bibitem{Courtois}
N.~T. Courtois, M.~Finiasz, and N.~Sendrier, ``How to achieve a
  {McEliece}-based digital signature scheme,'' in {\em International Conference
  on the Theory and Application of Cryptology and Information Security},
  pp.~157--174, Springer, 2001.

\bibitem{Stern}
J.~Stern, ``A method for finding codewords of small weight,'' in {\em
  International Colloquium on Coding Theory and Applications}, pp.~106--113,
  Springer, 1988.

\bibitem{Lee}
P.~J. Lee and E.~F. Brickell, ``An observation on the security of {McEliece's}
  public-key cryptosystem.,'' in {\em Eurocrypt}, vol.~88, pp.~275--280,
  Springer, 1988.

\bibitem{Baldi}
M.~Baldi, M.~Bodrato, and F.~Chiaraluce, ``A new analysis of the {McEliece}
  cryptosystem based on {QC-LDPC} codes,'' {\em Security and Cryptography for
  Networks}, pp.~246--262, 2008.

\bibitem{Bernstein}
D.~J. Bernstein, T.~Lange, and C.~Peters, ``Attacking and defending the
  {McEliece} cryptosystem,'' in {\em Post-Quantum Cryptography. PQCrypto 2008},
  pp.~31--46.

\bibitem{Li}
Y.~X. Li, R.~H. Deng, and X.~M. Wang, ``On the equivalence of {McEliece's} and
  {Niederreiter's} public-key cryptosystems,'' {\em IEEE Transactions on
  Information Theory}, vol.~40, no.~1, pp.~271--273, 1994.

\bibitem{Shor}
P.~W. Shor, ``Polynomial-time algorithms for prime factorization and discrete
  logarithms on a quantum computer,'' {\em SIAM review}, vol.~41, no.~2,
  pp.~303--332, 1999.

\bibitem{Shor2}
P.~W. Shor, ``Algorithms for quantum computation: Discrete logarithms and
  factoring,'' in {\em Foundations of Computer Science, 1994 Proceedings., 35th
  Annual Symposium on}, pp.~124--134, Ieee, 1994.

\bibitem{Deutsh}
D.~Deutsch and R.~Jozsa, ``Rapid solution of problems by quantum computation,''
  {\em Proc. R. Soc. Lond. A}, vol.~439, no.~1907, pp.~553--558, 1992.

\bibitem{Simon}
D.~R. Simon, ``On the power of quantum computation,'' {\em SIAM journal on
  computing}, vol.~26, no.~5, pp.~1474--1483, 1997.

\bibitem{NDavidMermin}
N.~D. Mermin, {\em Quantum computer science: an introduction}.
\newblock Cambridge University Press, 2007.

\bibitem{Kaye}
P.~Kaye, R.~Laflamme, and M.~Mosca, {\em An introduction to quantum computing}.
\newblock Oxford University Press, 2007.

\bibitem{Vazirani}
M.~Grigni, L.~Schulman, M.~Vazirani, and U.~Vazirani, ``Quantum mechanical
  algorithms for the nonabelian hidden subgroup problem,'' in {\em Proceedings
  of the thirty-third annual ACM symposium on Theory of computing}, pp.~68--74,
  ACM, 2001.

\bibitem{Lomont}
C.~Lomont, ``The hidden subgroup problem-review and open problems,'' {\em arXiv
  preprint quant-ph/0411037}, 2004.

\bibitem{Dinh}
H.~Dinh, C.~Moore, and A.~Russell, ``Mceliece and niederreiter cryptosystems
  that resist quantum fourier sampling attacks,'' in {\em Annual Cryptology
  Conference}, pp.~761--779, Springer, 2011.

\bibitem{Kempe}
J.~Kempe and A.~Shalev, ``The hidden subgroup problem and permutation group
  theory,'' in {\em Proceedings of the sixteenth annual ACM-SIAM symposium on
  Discrete algorithms}, pp.~1118--1125, Society for Industrial and Applied
  Mathematics, 2005.

\bibitem{Gulliver_thesis}
T.~A. Gulliver, {\em Construction of quasi-cyclic codes}.
\newblock PhD thesis, University of Victoria, 1989.

\bibitem{Aylaj}
B.~Aylaj, M.~Belkasmi, S.~Nouh, and H.~Zouaki, ``Good quasi-cyclic codes from
  circulant matrices concatenation using a heuristic model,'' {\em
  International journal of advanced computer science and applications}, vol.~7,
  no.~9, pp.~63--68, 2016.

\bibitem{Ling}
A.~Zeh and S.~Ling, ``Decoding of quasi-cyclic codes up to a new lower bound on
  the minimum distance,'' in {\em Information Theory (ISIT), 2014 IEEE
  International Symposium on}, pp.~2584--2588, IEEE, 2014.

\bibitem{Dixon}
J.~Dixon and B.~Mortimer, {\em Permutation Groups}.
\newblock Graduate Texts in Mathematics, Springer New York, 1996.

\bibitem{Our}
U.~{Kapshikar} and A.~{Mahalanobis}, ``{A Quantum-Secure Niederreiter
  Cryptosystem using Quasi-Cyclic Codes},'' {\em ArXiv e-prints}, Mar. 2018.

\bibitem{Hirotomo}
M.~Hirotomo, M.~Mohri, and M.~Morii, ``A probabilistic computation method for
  the weight distribution of low-density parity-check codes,'' in {\em
  International Symposium on Information Theory}, 2005.

\bibitem{NiedBook}
H.~Niederreiter and C.~Xing, {\em Algebraic Geometry in Coding Theory and
  Cryptography}.
\newblock Princeton University Press, 2009.

\end{thebibliography}
